\pdfoutput=1
\RequirePackage{amsmath} 
\documentclass[runningheads]{llncs}

\usepackage[T1]{fontenc}
\usepackage[utf8]{inputenc}
\usepackage{pslatex}
\usepackage{graphicx}
\usepackage{cite}               
\usepackage[hyphenbreaks]{breakurl}           
\usepackage[hyphens]{url}                
\usepackage[]{hyperref}
\usepackage{xcolor}             
\hypersetup{
  colorlinks,
  linkcolor={green!80!black},
  citecolor={red!70!black},
  urlcolor={blue!70!black}
}

\usepackage{algorithm} 
\usepackage[noend]{algpseudocode} 
\usepackage{amssymb} 
\usepackage{xspace}
\usepackage{balance}
\usepackage[inline]{enumitem} 
\usepackage{pifont}
\usepackage{epstopdf}
\usepackage{colortbl} 
\usepackage{mathtools} 
\usepackage{siunitx} 
\usepackage{import}
\usepackage{etoolbox}
\usepackage{framed} 
\patchcmd{\quote}{\rightmargin}{\leftmargin{} 1em \rightmargin}{}{}
\usepackage[latex]{svg} 
\usepackage{multicol} 
\usepackage{subcaption} 
\usepackage{listings} 
\usepackage{subfiles} 
\usepackage{titling}
\usepackage{booktabs} 
\usepackage{soul} 
\usepackage{local} 

\usepackage{tikz}
\usepackage{pgfplots}
\usepackage{pgfplotstable}

\usetikzlibrary{plotmarks} 
\pgfplotsset{compat=1.17}
\usepgfplotslibrary{groupplots} 
\usepgfplotslibrary{units} 

\lstdefinestyle{VisualStudio}{
  xleftmargin=0pt,
  basicstyle=\ttfamily\footnotesize,
  commentstyle=\color{CommentColor}\ttfamily\footnotesize,
  stringstyle=\color{darkred},
  keywordstyle=\color{BlackColor}\bfseries,
  escapeinside={(*@}{@*)}
}
\begin{document}

\date{}

\title{Kirigami, the Verifiable Art of Network Cutting}

\author{
{\rm Tim Alberdingk Thijm}\\
Princeton University
\and
{\rm Ryan Beckett}\\
Microsoft Research
\and
{\rm Aarti Gupta}\\
Princeton University
\and
{\rm David Walker}\\
Princeton University
}

\maketitle
%
\begin{abstract}
We introduce a modular verification approach to network control plane verification, where we cut a network into smaller fragments
to improve the scalability of SMT solving.
Users provide an annotated cut which describes how to generate these fragments from the monolithic network, and
we verify each fragment independently, using the annotations to define assumptions
and guarantees over fragments akin to assume-guarantee reasoning.
We prove this modular network verification procedure is sound and complete
with respect to verification over the monolithic network.
We implement this procedure as \sysname{}, an extension of \nv{}~\cite{giannarakis2020nv} --- a
network verification language and tool ---
and evaluate it on industrial topologies with synthesized policies.
We observe a 2--8x improvement in end-to-end \nv{} verification time,
with SMT solve time improving by up to 6 orders of magnitude.

\end{abstract}
%

\section{Introduction}\label{sec:intro}

Networks have become incredibly vast and labyrinthine systems.
To determine the best paths routers may use to forward traffic,
networks typically run distributed routing protocols.
Despite advances like software-defined networking,
these protocols remain widely used.
They are controlled by
millions of lines of decentralized, low-level router configuration code.
Operators must individually provision, maintain and reconfigure
the network's devices over time.
This overwhelmingly complexity has led to many notable outages~\cite{facebook-spineless,bgp-super-blunder,azure-outage},
with at times devastating pecuniary losses.
More often than not, the culprits behind these incidents
are subtle network misconfigurations.

In response, researchers have developed a variety of
verification tools and techniques
to catch errors before outages occur.
Some~\cite{anteater,hsa,veriflow,netplumber,netkat,nod,surgery,secguru}
have targeted the network \emph{data plane},
which is responsible for
forwarding traffic from point A to point B.
This work has produced scalable and performant methods for modeling the
data plane and checking properties of how packets traverse it.

The data plane is produced by the network's
\emph{control plane}, which uses the aforementioned
routing protocols to decide which routes to use.
Occasionally,
these protocols may update their choice of routes ---
\EG{} following a device failure ---
and recompute new paths.
When this happens, the data plane is
regenerated, and the user must repeat any data plane analysis.
Obscure control plane faults can lead to further
issues,
and manual verification by a human operator
is an effort in locating what may be a minuscule typo
within a gargantuan morass of router configurations.

To address this problem, researchers have developed
another suite of tools to analyze the control plane~\cite{batfish,minesweeper,arc,era,bagpipe,bonsai,origami,shapeshifter,tiramisu,giannarakis2020nv}.
Control plane analyses consider what routes will be used by the
data plane in given network environments,
and check properties of the network in such environments.
These tools can uncover bugs
in real networks, but unfortunately tend not to scale
as well as their data plane counterparts.

One branch of control plane verification, starting from Minesweeper~\cite{minesweeper},
encodes a network as a Satisfiability Modulo Theories (SMT) formula and then
asks an SMT solver~\cite{smt} to check properties of the encoded network.
While SMT-based verification has advantages over other approaches, such as its expressivity
or applications in automated network repair~\cite{cpr},
it nonetheless suffers from scalability issues.
Prior work has explored using abstractions to resolve this problem, \EG
using symmetries in network topologies to compress networks~\cite{bonsai,origami}.
These abstractions offer some relief, but cannot always handle arbitrary non-symmetrical networks.

This paper offers another path forward in scaling SMT-based control plane
verification, by being the first to leverage the inherent modularity
of the control plane to \emph{cut} a monolithic network
into multiple fragments to be verified independently.
Building on prior work on assume-guarantee verification
of modular programs~\cite{henzinger1998you,giannakopoulou2018compositional},
we present a novel technique for modular verification of control planes
and implement it as \sysname{}, an extension to the \nv{}~\cite{giannarakis2020nv} network verification
language and tool.

In a typical assume-guarantee verification approach,
one can verify a safety property $P$ over a system of concurrent processes, by verifying 
each process
independently,
using \emph{assumptions} over the environment in which it runs and \emph{guarantees} over how it modifies this environment.
The relationships between assumptions and guarantees (formulated as assume-guarantee rules) are then checked, which allows one to
conclude that if all checks pass,
then $P$ holds for the monolithic system.
Our verification technique mirrors this idea: we verify a property over fragments (\CF processes) of the control plane,
given assumptions over the rest of the network and guarantees over our fragments,
to conclude that the monolithic network respects the property.

We start from an existing general model for distributed routing,
the Stable Routing Problem (SRP) model~\cite{bonsai}.
In an SRP, each node of the network exchanges routes
with its neighbors to compute a locally-stable solution.
Like other work in control plane verification~\cite{arc,tiramisu,fastplane,shapeshifter},
we focus on networks (\IE SRPs) with unique solutions.
We develop an SRP extension called ``open SRPs'', in which
a network receives routes along a set of \emph{input nodes}
and sends out routes along a different set of \emph{output nodes}.
We identify the solutions of our input nodes as our
open SRP's assumptions, and the solutions of our output nodes
as its guarantees.
We present a procedure $\cut$ which,
given an \emph{interface} --- a mapping from a cut-set of edges to routes ---
cuts an open SRP $S$ into two
smaller open SRPs $T_{1}$ and $T_{2}$ covering $S$, and where each cut edge
is replaced by a route assumed in one SRP and guaranteed in the other.
Interfaces can follow a network's natural boundaries,
\EG{} a data center network interface might be cut according to its levels or hierarchy~\cite{fattree,dcell,bcube}.

As with the traditional (closed) SRP, we can check that an open SRP
satisfies a given safety property $P$ by verifying that $P$ holds
for the SRP's solutions.
We prove that if $P$ holds on $T_{1}$ and $T_{2}$'s
solutions, then it holds on $S$'s.
This is the basis for our modular network verification technique.
Starting from a network $S$, an interface $I$, and a safety property $P$,
we use $\cut(S,I)$ to obtain a set
of $N$ open SRPs $T_{1}, \ldots, T_{N}$ that are verified
independently.
We verify $P$ and $T_{i}$'s guarantees for each open SRP $T_{i}$:
if either the property or interface's guarantees do not hold,
we return a counterexample demonstrating the solution that does not satisfy $P$ or $I$.
We believe this to be the first work to present a proven-correct general theory for
automated modular verification of arbitrary properties, and where we check the correctness
of the given interface:
prior work on modular network verification like~\cite{secguru} considered specific architectures
and properties without any guarantee of correctness.

As SMT-based verification time typically grows
superlinearly with the size of the network~\cite{minesweeper},
by verifying $P$ on each of the $N$ smaller open SRPs $T_{i}$,
we can verify $P$ in a fraction of the time it takes
to do so directly over the monolithic network $S$.
Our experiments demonstrate that this modular verification technique works well for a variety of data center,
random and backbone networks, with significant improvements in SMT solve time:
we show for one set of fattree~\cite{fattree} benchmarks that verifying the fattree pod-by-pod cuts SMT time
from 90 minutes to under 2 seconds; verifying every node individually reduces SMT time to around
a hundredth of a second.
Overall, while we are working on improving the engineering in \nv{} for carving out partitions, 
we already see a 2--8x speedup in end-to-end \nv{} verification time.
We also observe that a modular approach assists in producing more localized errors
and debugging feedback in the cases when verification fails.

In summary, we make the following contributions:
\begin{description}
  \item[A Theory of Network Fragments]
        We develop an extension of the Stable Routing Problem (SRP) model~\cite{bonsai} for network fragments.
        Our extension provides a method to cut monolithic SRPs into a set of fragments.
        We define \emph{interfaces} to cut SRPs and map the cut edges to annotations which then define
        \emph{assumptions} and \emph{guarantees} of our fragments.
        We prove that under these assumptions, if these guarantees hold, then
        a property that holds in every fragment also holds in the monolithic network.
  \item[A Modular Network Verification Technique]
        We present a technique to decompose a monolithic network verification problem into multiple subproblems.
        We start from an SRP $S$, an interface $I$ and a property $P$, and cut $S$ into a set of fragments.
        We check each fragment's guarantees and the given $P$
        independently and report whether $P$ holds for $S$, or if $P$ or $I$ fail to hold.
        This enables a novel, modular approach to control plane verification 
        based on assume-guarantee reasoning.
  \item[Fast, Scalable and Modular SMT Verification]
        We implement \sysname, a tool based on this theory, as an extension for \nv{},
        a network verification language and tool~\cite{giannarakis2020nv}.
        Using \sysname, we improve on \nv{} verification in terms of scalability and performance.
        SMT solve time using \sysname is up to \emph{six orders of magnitude} faster for a selection of \nv benchmarks.
\end{description}

\section{Overview}\label{sec:overview}

\para{The Stable Routing Problem}
A network is a graph with nodes $V$ representing routers and edges
$E$ representing the links between them.
A distributed control plane uses routing protocols to determine paths to routing destinations.
Each router deploys its own local rules to broadcast routing announcements (or \emph{routes})
and select a ``best'' route: the form of these rules varies with the protocol,
but generally protocols focus on minimizing routing costs.

These elements --- nodes and edges, a set of routes,
and a set of rules to initialize, compare and broadcast them ---
form the basis for our
control plane routing model, the SRP~\cite{bonsai}\@.
In a well-designed network, this exchange of routes eventually converges to a \emph{stable state},
where no node may improve on its current best route by selecting another offered by a neighbor.
A mapping from nodes to these stable routes is called a \emph{solution} $\lab$ to the SRP\@.
While it is possible for routing to diverge (\IE{} have no solution) or converge to multiple solutions,
many typical networks have unique solutions (\EG when routing costs strictly increase with distance to the destination~\cite{arc,fastplane}): we restrict our focus in this paper to
such networks, like other work~\cite{arc,fastplane,tiramisu,shapeshifter}.

\begin{figure*}[t]
  \centering
  \begin{subfigure}{0.45\linewidth}
    \includesvg[pretex=\small,width=1.0\linewidth]{img/fat20}
    \caption{\label{fig:fat20-topology} A fattree topology.}
  \end{subfigure}
  \begin{subfigure}{0.45\linewidth}
\begin{lstlisting}[linerange={1-2,5-7,14-19}]
type attribute = int (*@\label{lst:fat-attrib}@*)

let nodes = 20 (*@\label{lst:fat-topology}@*)(* topology *)
let edges = {
  0=4; 0=6; 0=8; 0=10;(*...*)
  10=18; 10=19; 11=18; 11=19;
}

let merge node x y = if x < y then x else y
let trans edge x = x + 1
let init node = if node = 19n then 0 else NULL
\end{lstlisting}
    \caption{\label{fig:fat20-nv} An \nv{} program \texttt{fat.nv} representing Figure~\ref{fig:fat20-topology}.}
  \end{subfigure}
  \begin{subfigure}{1.0\linewidth}
\begin{lstlisting}
include "fat.nv"

(* map each node to its solution (stable route) *)
let sol = solution { init = init; trans = trans; merge = merge; }(*@\label{lst:fat-sol}@*)
(* check a property [route <= 4] of every node's solution *)
assert foldNodes (fun node route acc -> acc && route <= 4) sol true (*@\label{lst:fat-assert}@*)
\end{lstlisting}
    \caption{\label{fig:fatnv-mono} An \nv{} program asserting that every node can reach \texttt{19n} in at most 4 hops.}
  \end{subfigure}
  \caption{A fattree network $S$ and its representation in \nv{}. Node $d$ (\texttt{19n} in \nv{})\protect\footnotemark%
   propagates an initial route to itself to the rest of the network.}
  \label{fig:fat20}
\end{figure*}
\footnotetext{\nv{} numbers nodes starting from \texttt{0n}: our example numbers
  Figure~\ref{fig:fat20-topology} from left to right,
  \IE \texttt{0n} is $c_{0}$, \texttt{4n} is $a_{0}$, \texttt{12n} is $e_{0}$ and so on.
``\texttt{0=4}'' refers to a bidirectional edge between \texttt{0n} and \texttt{4n}.}

\para{An Example SRP}
Let's consider an SRP instance $S$ of a familiar fattree~\cite{fattree} data center network,
as shown in Figure~\ref{fig:fat20-topology}.
Routing in fattree networks typically follows a $\Lambda$ shape:
traffic that starts from an edge switch ($e_{0}, \ldots, e_{6}, d$) travels up along a link
to an aggregation switch ($a_{0}, \ldots, a_{7}$), then ascends from the pod to a core switch ($c_{0}, \ldots, c_{3}$)
in the spine
before descending back down into another pod.
For this example, our routes will simply be the number of hops to some
routing destination $d$. Initially $d$ will know a route with 0 hops to itself;
the rest of the network starts with no route to $d$.
Each node broadcasts its route to $d$ to all of its neighbors,
incrementing the route by one hop.
Nodes will then compare each received route with their current choice and select the
one with the fewest hops.
The unique solution $\compof{S}{\lab}(u)$ of a node $u$ in $S$
is thus the best route between $u$'s initial route and the transferred solutions of each of $u$'s neighbors. 
This toy policy elides the complexities of real routing protocols, which may have dozens of fields,
each with particular semantics, but demonstrates all the basic elements of an SRP.

\para{Verifying SRPs with \nv{}~\cite{giannarakis2020nv}}
We can verify properties of $S$'s solution to confirm our beliefs about $S$'s behavior.
For instance, we may wish to check that every node's route to $d$ is at most 4 hops.
One verification tool we can use to do so is \nv{}~\cite{giannarakis2020nv}.
\nv{} is a functional programming language for modeling control planes
with an associated SMT verification engine.
An \nv{} program's components map onto those of an SRP:
it has a topology (\texttt{nodes} and \texttt{edges}); a type of routes (\texttt{attribute});
a function \texttt{init} to initialize routes; a function \texttt{trans} to broadcast routes;
and finally a function \texttt{merge} to compare routes.
Figure~\ref{fig:fat20-nv} presents a condensed \nv{} program for Figure~\ref{fig:fat20-topology}.

Figure~\ref{fig:fatnv-mono} demonstrates how to verify a safety property $P$ in \nv{},
where $P$ holds \IFF $\forall u.~\lab(u) \leq 4$.
We define the solution (line~\ref{lst:fat-sol})
using \texttt{init}, \texttt{trans} and \texttt{merge} from Figure~\ref{fig:fat20-nv}.
We then assert (line~\ref{lst:fat-assert}) that $P$ is true of this solution.
When we supply Figure~\ref{fig:fatnv-mono} to \nv{}'s verification engine, \nv{} encodes $S$
and $P$ as an SMT query, and confirms that $P$ holds for $\compof{S}{\lab}$.
Encoding the network to SMT lets us reason about network states symbolically,
avoiding state explosion when analyzing properties like fault tolerance or reasoning about routes arriving from
outside the network.

\para{Scaling Up SRP Verification}
SMT-based verification is expressive, but has issues when it comes to scalability.
Our evaluation in \S\ref{sec:evaluation} shows that SMT verification scales superlinearly for larger fattrees with
more complex policies:
from 0.03 seconds for a 20-node network, to 1.41 seconds for an 80-node network, and 1833.66 seconds for a 320-node network!
To verify the tens of thousands of switches in industrial fattree networks~\cite{secguru}, we must find a way to scale this
technique.

Suppose then that we took a large network and \emph{cut it into fragments} (defined formally in \S\ref{sec:partitions}),
in order to verify a safety property $P$ on each fragment independently.
In other words, if $P$ holds for every node in every fragment,
then it holds for every node in the monolithic network;
and otherwise, we want to observe real counterexamples as in the monolithic network.
To achieve this goal, our cutting procedure must also summarize the network behavior external to each fragment.

We incorporate these summaries into the traditional SRP model by generalizing it to open SRPs.
Open SRPs extend the SRP model by designating some nodes as \emph{input nodes} and some others as \emph{output nodes}.
Input and output nodes are annotated with routes representing solutions \emph{assumed} on the inputs
and \emph{guaranteed} on the outputs.
We express these annotations using an \emph{interface}:
a mapping from each cut edge to a route annotation.
Given an open SRP $S$ and an interface $I$,
we cut $S$ into open SRP fragments, where each fragment identifies
assumptions on its inputs and guarantees on its outputs.

\para{Cutting Down Fattrees}
We will now move on to demonstrating this idea for Figure~\ref{fig:fat20}.
Let's cut each pod of our network into its own fragment $T_{p0}$ through $T_{p3}$,
leaving the spine nodes as a fifth fragment $T_{spines}$.
\begin{figure}[t]
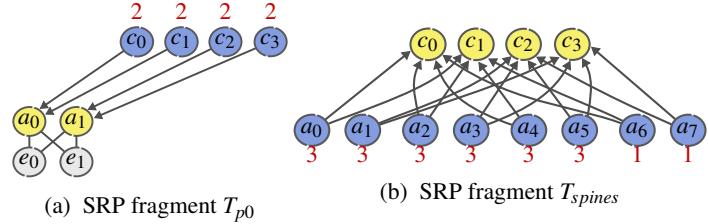

  \centering
  \begin{subfigure}{0.3\linewidth}
    \includesvg[pretex=\footnotesize,width=1.0\linewidth]{img/fat20-pod0}
    \caption{\label{fig:fat20frag0} SRP fragment $T_{p0}$}
  \end{subfigure}
  \begin{subfigure}{0.45\linewidth}
    \includesvg[pretex=\footnotesize,width=1.0\linewidth]{img/fat20-spines}
    \caption{\label{fig:fat20frag4} SRP fragment $T_{spines}$}
  \end{subfigure}
  \caption{\label{fig:fat20frags} SRP fragments $T_{p0}$ and $T_{spines}$,
    with input nodes in blue, output nodes in yellow and
    assumptions written in red.}
\end{figure}
Figures~\ref{fig:fat20frags}\subref{fig:fat20frag0} and~\ref{fig:fat20frags}\subref{fig:fat20frag4} show pod 0
and the spines of Figure~\ref{fig:fat20} as open SRPs $T_{p0}$ and $T_{spines}$, respectively.
In $T_{p0}$, we assume routes from the spines and check guarantees on $a_{0}$ and $a_{1}$.
An assumption in one fragment will be guaranteed by another (and vice-versa):
we assume $a_{0}$ has a route of 3 hops in $T_{spines}$ and check that it has a route of 3 hops in $T_{p0}$.

\para{Verifying Network Fragments}
In modular verification,
we perform an independent verification query for each fragment:
we encode the open SRP and property, along
with an assumptions formula assuming a state of the inputs
and a guarantees formula to check on the state of the outputs.
We then submit every query to our solver and ask if the network has a solution where,
under the given assumptions,
either the property is false (as before) or the guarantees formula does not hold.
Our solver then searches for a counterexample demonstrating a concrete violation
of the property or our guarantees.
Guarantee violations provide evidence of possible bugs in our network implementation or mistakes in our beliefs,
in the same way that property violations do.

Let us consider our fattree network again.
Suppose we misconfigured $a_{6}$ to black hole (silently drop) traffic, leading
nodes $a_{0}$, $a_{2}$ and $a_{4}$ to re-route via the other nodes $a_{1}, a_{3}, a_{5}$ in their respective pods.
Our interface maps $c_{0}a_{0}$ to $2$,
so we check that $\compof{spines}{\lab}(c_{0}) = 2$ when verifying $T_{spines}$.
Due to our bug, this check fails and our solver returns
a counterexample: because $c_{0}$ must reroute, $\compof{spines}{\lab}(c_{0}) = 6$.
We can then modify our network configuration to fix the bug, and re-run verification to confirm
that our guarantees and property hold for all fragments.
We prove in \S\ref{sec:partitions} that this implies that $P$ holds for the monolithic network.
Our guarantees can thus be thought of as a \emph{specification} of the
desired network behavior along its cut points, in addition to $P$.

\begin{figure}[!t]
  \centering
\begin{lstlisting}[linerange={1-4,8-16}]
include "fat.nv"

let partition node = match node with
  | 0n | 1n | 2n | 3n -> 0 (* spines *)(*...*)
  | 10n | 11n | 18n | 19n -> 4 (* p3 *)

let interface edge x = match edge with
  | 0~_ | 1~_ | 2~_ | 3~_ -> x = 2
  | 4~_ | 5~_ | 6~_ | 7~_ | 8~_ | 9~_ -> x = 3
  | 10~_ | 11~_ -> x = 1

let sol = solution { init = init; trans = trans; merge = merge; interface = interface }
assert foldNodes (fun node route acc -> acc && route <= 4) sol true
\end{lstlisting}
  \caption{An \nv{} program which cuts Figure~\ref{fig:fat20} into pods
    (some node and edge cases not shown).}
  \label{fig:fatnv-pods}
\end{figure}

\para{Verifiable Network Cutting with \sysname{}}
As part of our work, we implemented an extension \sysname{} to \nv{}
for cutting and verifying networks.
Figure~\ref{fig:fatnv-pods} shows an \nv{} file
with two new functions, \texttt{partition} and \texttt{interface}.
\texttt{partition} assigns each node to a fragment, while \texttt{interface} adds assertions that check that
a route $x$ along a cross-fragment edge is equal to the specified annotation,
\EG that the route from \texttt{0n} to \texttt{4n} is 2 hops.
Under the hood, we cut the network using \texttt{partition} to generate our fragments,
and then annotate the cuts using \texttt{interface}; verification can then proceed
as described.

\para{A Cut Above the Rest}
Pod-based cuts suit our high-level understanding of fattrees, but
we can consider many other cuts.
We could cut Figure~\ref{fig:fat20} so that
every node is in its own fragment.
Verifying a single node in SMT is extremely cheap, and hence
leads to significant performance improvements.
The corresponding \nv{} program resembles Figure~\ref{fig:fatnv-pods},
except every node maps to its own fragment and we annotate every edge.

\para{Next Steps}
The rest of the paper proceeds as follows.
\S\ref{sec:srp} presents prior work formalizing SRPs, and
\S\ref{sec:partitions} presents our extensions for cutting SRPs,
with proofs of soundness and completeness of our procedure.
We present our SMT checking procedure in \S\ref{sec:algorithm}, and
the implementation of our theory in \S\ref{sec:implementation} as
\sysname{}, an extension of \nv{}.
We evaluate \sysname{} in \S\ref{sec:evaluation}.
We discuss related work in \S\ref{sec:related},
and future work in \S\ref{sec:discussion}.

\section{Background on the Stable Routing Problem}\label{sec:srp}

We summarize prior work~\cite{bonsai} on the Stable Routing Problem (SRP) network model.
Many components of this model resemble routing algebras used for reasoning about convergence of routing protocols~\cite{routingalgebra,metarouting,daggitt2018asynchronous}, but SRPs also include a network topology
for reasoning about properties such as reachability between nodes.

An SRP instance $S$ is a 6-tuple $(V, E, R, \init, \merge, \transfer)$, defined as follows.

\para{Topology}
$V$ is a set of nodes and $E \subseteq V \times V$ is a set of directed edges between them.
We write $uv$ for an edge from node $u$ to node $v$.
Edges may not be self-loops: $\forall v \in V.~vv \notin E$.

\para{Routes}
$R$ is a set of routes that describe the fields of routing messages.
For example, when modeling BGP, $R$ might represent a tuple of an integer local preference,
a set of community tags, and a sequence of AS numbers representing the AS path~\cite{RFC1997,RFC4271}.

\para{Node Initialization}
The initialization function $\init : V \rightarrow R$ describes the initial route of each node.
When modeling single destination routing, $\init$ may map a destination node $d$ to some initial route $r_{d}$,
and all other nodes to a null route; in multiple destination routing, we may have many initial routes.

\para{Route Update}
The merge function $\merge : R \times R \rightarrow R$ defines how to compare and merge routes.
$\merge$ represents updates of a node's selected route: we assume $\merge$ is associative
and commutative, \IE the order in which a sequence of routes are merged does not matter.

\para{Route Transfer}
The transfer function $\transfer : E \times R \rightarrow R$ describes how routes are modified between nodes.
Given an edge $uv$ and a route $r$ from node $u$, $\transfer(uv,r)$ determines the route received at $v$.

\para{Solutions}
A solution $\lab : V \rightarrow R$ is a mapping from nodes to routes.
Intuitively, a solution is defined such that each node is \emph{locally stable},
\IE it has no incentive to deviate from its currently chosen neighbors.
Nodes compute their solution via message exchange, where each node in the SRP advertises
its chosen route to each of its neighbors.
Formally, an SRP solution $\lab$ satisfies the constraint:
\begin{equation}\label{eq:srp-sol}
  \lab(v) = \init(v) \merge \Merge_{uv \in E} \transfer(uv, \lab(u))
\end{equation}
\noindent where $\Merge$ is the sequence of $\merge$ operations
on each transferred route $\transfer(uv, \lab(u))$ from each neighbor $u$ of $v$.
These received routes are merged with $v$'s initial value $\init(v)$.

A solution may determine an SRP's forwarding behavior or another decision-making procedure,
as shown in~\cite{bonsai}.
We omit discussing forwarding behavior in this work to focus on a general SRP definition
without restricting ourselves only to forwarding.

\section{Cutting SRPs}\label{sec:partitions}

We now introduce our original contributions, starting with \emph{open SRPs}.
We define a fragment relation between a smaller open SRP and a larger one,
and define a $\cut$ procedure to decompose one open SRP into a partition of two fragments.
We prove soundness and completeness of partition solutions with respect to the larger SRP's solution.

\para{Notation}
We introduce some notation in this section that may be unfamiliar.
$\dom(f)$ is the \emph{domain} of the function $f$, and
$f |_{X}$ is the \emph{restriction} of $f$ to $X \subseteq \dom(f)$.
We use subscripts to specify SRP components, \EG
$\compof{S}{\init}$ refers to SRP $S$'s $\init$ component.

\begin{figure}[t]
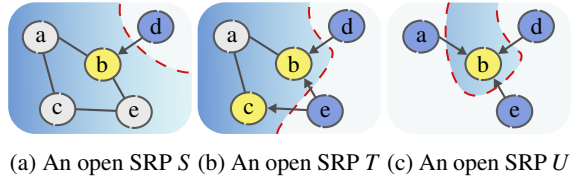

  \centering
  \begin{subfigure}{0.2\linewidth}
    \includesvg[width=\linewidth]{img/cut-d}
    \caption[SRP S]{An open SRP $S$}\label{fig:srp-s}
  \end{subfigure}
  \begin{subfigure}{0.2\linewidth}
    \includesvg[width=\linewidth]{img/cut-de}
    \caption[SRP T]{An open SRP $T$}\label{fig:srp-t}
  \end{subfigure}
  \begin{subfigure}{0.2\linewidth}
    \includesvg[width=\linewidth]{img/cut-deca-no-shadow}
    \caption[SRP U]{An open SRP $U$}\label{fig:srp-u}
  \end{subfigure}
  \caption{A series of successive cuts which produce open SRPs $S$, $T$ and $U$.
    Base nodes are shown in grey, input nodes in blue and output nodes in yellow.
    Cuts between input and non-input nodes are shown in red.}
  \label{fig:fragments}
\end{figure}

\para{Open SRPs}
An \emph{open SRP} generalizes our earlier SRP definition to include
\emph{assumptions} and \emph{guarantees}.
An open SRP instance $S$ is an 8-tuple $(V, E, R, \init, \merge, \transfer, \inh, \outh)$.

The first six elements are defined exactly as for regular (closed) SRPs.
The final two elements, $\inh$ (``assumptions'') and $\outh$ (``guarantees''),
are \emph{partial functions} ($V \hookrightarrow R$)
mapping mutually disjoints subsets $\vin, \vout \subseteq V$ to routes.
We use $\vin$ (input nodes) as a shorthand for $\dom(\inh)$
and $\vout$ (output nodes) as a shorthand for $\dom(\outh)$.
All nodes that are neither input nor output nodes are ``\emph{base nodes}'' $\basenodes$.
A closed SRP is an open SRP where $\vin = \vout = \emptyset$.
Going forward, we assume an open SRP whenever we write ``SRP'',
except when the distinction is relevant.

Input nodes must be \emph{source nodes} (in-degree $= 0$).
Hence, they act as auxiliary nodes, indicating where a fixed incoming
route ``arrives'' from outside the SRP, as specified by the assumptions $\inh$.
Output nodes correspondingly mark where routes ``depart'' the SRP, per the guarantees $\outh$.
We do not require any connectivity properties of output nodes:
we think of them as simply identifying an outgoing route we wish to guarantee, but
without requiring the SRP to tell us whither it is announcing that route.%
\footnote{This is partly a design choice: we could have also attached auxiliary nodes
  to output nodes to indicate where these routes are going, but we found this definition the most
straightforward.}
Figure~\ref{fig:fragments} illustrates this concept with some example open SRPs and cuts.

\begin{definition}[Open SRPs]\label{defn:open-srp}
  An \emph{open SRP instance} $S = (V, E, R, \init, \merge, \transfer, \inh, \outh)$
  respects the following properties:
  \begin{itemize}
    \item $V = \vin \cup \vout \cup \basenodes$ and $\vin, \vout, \basenodes$ are pairwise-disjoint;
    \item $\inh : \vin \rightarrow R$ and $\outh : \vout \rightarrow R$; and
    \item $\forall v \in \vin.~\text{in-degree}(v) = 0$.
  \end{itemize}
\end{definition}

\para{Open SRP Solutions}\label{sec:open-sol}
A mapping $\lab$ is a \emph{solution} to an open SRP \IFF:
\begin{align}
  \lab(u) &=
    \init(u) \merge \Merge_{vu \in E}\transfer(vu, \lab(v)) & \forall v &\notin \vin \label{eq:open-sol}\\
  \lab(u) &= \inh(u) &    \forall v &\in \vin \label{eq:open-sol-in}\\
  \lab(u) &= \outh(u) &   \forall v &\in \vout \label{eq:open-sol-out}
\end{align}

\noindent Note that Equations~\eqref{eq:open-sol} and~\eqref{eq:open-sol-out} both apply for all outputs $v \in \vout$.
Solutions for open SRPs resemble closed SRP solutions, with the addition of constraints
based on the values of $\inh$ and $\outh$.
For any input node $u$, its assumption $\inh(u)$ determines the node's solution directly;
for an output node $u$, its solution $\lab(u)$ must be consistent with both
the right-hand side of~\eqref{eq:open-sol} \emph{and} the right-hand side of~\eqref{eq:open-sol-out}.
Hence, if $\exists u \in \vout.~\init(u) \merge \Merge_{vu \in E}\transfer(vu, \lab(v)) \not= \outh(u)$,
there is no solution to the open SRP.
As with closed SRPs, we restrict our focus to open SRPs with unique solutions.

\para{Fragments}\label{sec:fragment-relation}
We now introduce a fragment relation between two open SRPs.
We may think of an open SRP as composed of fragments of smaller open SRPs,
where each fragment represents its connection to the rest of the larger SRP
with assumptions and guarantees.
Consider the series of open SRPs in Figure~\ref{fig:fragments}.
One can think of $T$ (Figure~\ref{fig:srp-t}) as a fragment representing part of $S$ (Figure~\ref{fig:srp-s}).
To go from $S$ to $T$, we can cut $e$ off from $S$ to obtain a network with
new input nodes which summarize the rest of the network with assumptions.
One can fragment $T$ further by cutting off $b$ from the remaining nodes:
this produces an even smaller fragment $U$ (Figure~\ref{fig:srp-u}).

\begin{definition}[Fragments]\label{defn:fragment}
  Let $S$ and $T$ be open SRPs.
  $T$ is a \emph{fragment} of $S$ when:
  \begin{align}
    \compof{T}{V} &\subseteq \compof{S}{V} 
    &\compof{T}{E} &= \{ uv ~\vert~ u \in \compof{T}{V}, v \in \compof{T}{V},
      uv \in \compof{S}{E}, v \notin \compof{T}{\vin} \} &\label{eq:frag-e} \\
    \compof{T}{R} &= \compof{S}{R} 
    &\infixcompof{T}{\merge} &= \infixcompof{S}{\merge} &\label{eq:frag-merge} \\
    \compof{T}{\init} &= \compof{S}{\init} |_{\compof{T}{V}} 
    &\compof{T}{\transfer} &= \compof{S}{\transfer} |_{\compof{T}{E}} &\label{eq:frag-transfer}
  \end{align}
  \begin{align}
    &\compof{T}{\vin} = (\compof{S}{\vin} \cup
                       \{ v ~\vert~ uv \in \compof{S}{E}, u \notin \compof{T}{V} \})
    \cap \compof{T}{V} &\label{eq:frag-inputs} \\
    &\compof{T}{\vout} = (\compof{S}{\vout} \setminus \compof{T}{\vin} \cup \{ u ~\vert~ uv \in \compof{S}{E}, v \notin \compof{T}{V} \}) \cap \compof{T}{V} &\label{eq:frag-outputs} \\
    &\forall u \in (\compof{T}{\vin} \cap \compof{S}{\vin}).~\compof{T}{\inh}(u) = \compof{S}{\inh}(u) &\label{eq:frag-ass} \\
    &\forall u \in (\compof{T}{\vout} \cap \compof{S}{\vout}).~\compof{T}{\outh}(u) = \compof{S}{\outh}(u) &\label{eq:frag-guar}
  \end{align}
\end{definition}

Informally, the fragment $T$ is made up of a subgraph of $S$ over nodes $\compof{T}{V}$,
conserving all edges from $\compof{S}{E}$ between them, except any edges into input nodes~\eqref{eq:frag-e}.
Routing and routing functions of $S$ are as before~\eqref{eq:frag-merge} or restricted over $T$'s topology~\eqref{eq:frag-transfer}.
Finally, $T$ designates nodes whose neighbors have been cut as inputs~\eqref{eq:frag-inputs} (summarizing
the network ``outside'' $T$) or outputs~\eqref{eq:frag-outputs} (communicating a summary to the ``external'' network),
while preserving any assumptions~\eqref{eq:frag-ass} and guarantees~\eqref{eq:frag-guar}
inherited from $S$.

\para{Interfaces and Cutting SRPs}\label{sec:interfaces}
The fragment relation leaves unspecified how a smaller SRP's
assumptions and guarantees summarize its parent's routes.
We now consider how to cut an SRP $S$ into two fragments $T_{1}$ and $T_{2}$,
where $T_{1}$ and $T_{2}$ cover $S$ and replicate its behavior with the
help of their assumptions and guarantees.
We do so by selecting a cut-set $C \subseteq E$ of edges in $S$
and annotating each cut edge $uv$ with a route that describes the
solution transferred from $u$ to $v$.
We call this annotated cut-set an \emph{interface} $I$.

\begin{definition}[Interface]\label{defn:interface}
  Let $S$ be an SRP and let $C \subseteq E$ be a cut-set partitioning $\compof{S}{V}$.
  $I : C \rightarrow \compof{S}{R}$ is an \emph{interface} if it maps
  every element $uv$ of $C$ to a route $I(uv)$ in $\compof{S}{R}$.
\end{definition}

We now define a $\cut$ procedure.
Given an SRP $S$ and an interface $I$,
$\cut(S, I)$ returns a \emph{partition} of two SRP fragments, $T_{1}$ and $T_{2}$.
Nodes along the cut edges are annotated with assumptions and guarantees.
We can recursively $\cut$ an SRP into arbitrarily many fragments.
We elide the structural details of how $\cut$ divides the nodes of $S$
between $T_{1}$ and $T_{2}$ for now: curious readers should see Appendix~\ref{sec:proofs}.

\begin{figure}[t]
  \centering
    \includesvg[width=0.7\linewidth]{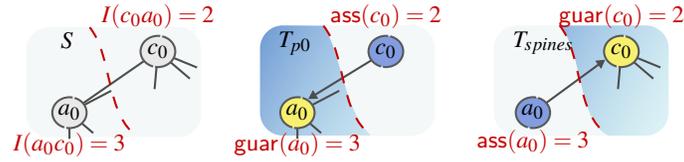}
  \caption{A closeup of how $\cut$ defines input-output nodes $c_{0}$ and $a_{0}$ from $I$.}
  \label{fig:fat-c0a0}
\end{figure}

What is most important about $\cut$ is that it defines
$T_{1}$ and $T_{2}$ to have \emph{equal} assumptions and guarantees
along each cut edge.
For each edge $uv$ in our interface $I$,
$\cut(S, I)$ adds a guarantee $\outh(u) = I(uv)$ in $T_{1}$
and an assumption $\inh(u) = I(uv)$ in $T_{2}$ (or vice-versa).
By requiring this equality, we rely on the stability of an open SRP's solution
to avoid the issue of circularity in assumptions.
The edge $uv$ now shows a route $I(uv)$ ``arriving'' in $T_{2}$
after ``departing'' from $T_{1}$.
As $u$'s solution is both assumed in one fragment and guaranteed in the other,
we refer to it as an \emph{input-output node}.
We illustrate this idea in Figure~\ref{fig:fat-c0a0}, which shows how an interface defines
assumptions and guarantees for input-output nodes $c_{0}$ and $a_{0}$ from Figure~\ref{fig:fat20frags}.

\begin{definition}[Input-output nodes]
  Let $T_{1}$ and $T_{2}$ be two open SRPs with a set
  $(\compof{1}{\vin} \cap \compof{2}{\vout}) \cup (\compof{2}{\vin} \cap \compof{1}{\vout})$
  of shared nodes.
  A node $u$ in this set is an \emph{input-output node} \IFF
  $\compof{1}{\inh}(u) = \compof{2}{\outh}(u)$ or $\compof{2}{\inh}(u) = \compof{1}{\outh}(u)$.
\end{definition}

Because $\cut$ produces input-output nodes, we can
reason over the solutions of both fragments separately, using the assumptions
and guarantees of our input-output nodes to confirm that the solutions coincide along our cut.
We can now define a \emph{partition} as a relation between $T_{1}$ and $T_{2}$ and $S$.

\begin{definition}[Partition]\label{defn:partition}
  Let $S$, $T_{1}$ and $T_{2}$ be open SRPs.
  $(T_{1}, T_{2})$ is a \emph{partition} of $S$ when
  \begin{enumerate*}[label=\emph{(\roman*)}]
    \item $T_{1}$ and $T_{2}$ are both fragments of $S$,
    \item $\compof{1}{V} \cup \compof{2}{V} = \compof{S}{V}$ and $\compof{1}{E} \cup \compof{2}{E} = \compof{S}{E}$,
    \item every input node in $T_{1}$ or $T_{2}$ that is not an input node in $S$ is an input-output node.
  \end{enumerate*}
\end{definition}

We present the full definition of a partition --- which includes some corner cases for
when two fragments share the same input node --- in Appendix~\ref{sec:proofs}:
these details are not required to understand our theorems.
We prove that our $\cut$ procedure always produces a partition, and subsequently
prove that if $T_{1}$ and $T_{2}$ are a partition of $S$, then the joined solutions of $T_{1}$ and $T_{2}$
are a solution of $S$ (soundness); and that if $S$ has a solution, then there always exists
an interface $I$ that given to $\cut$ produces a partition of two fragments $T_{1}$ and $T_{2}$
such that the solution of $S$ is a solution (when appropriately restricted) for $T_{1}$ and $T_{2}$ (completeness).

\begin{definition}[$\cut$]\label{defn:cut}
  Let $S$ be an SRP and let $I$ be an interface over $S$.
  Given $S$ and $I$, $\cut(S, I) = (T_{1}, T_{2})$,
  where $T_{1}$ and $T_{2}$ are a \emph{partition} of $S$ such that
  $\forall uv \in \dom(I)$, $u$ is an input-output node between $T_{1}$ and $T_{2}$.
\end{definition}

\para{Correctness}\label{sec:partition-theorems}
We now prove theorems on the relationships between an SRP's solution
and the solutions of its $\cut$-produced fragments.
By showing that the fragments' solutions are the same as the monolithic SRP's,
we can use the fragments \emph{in place of} the monolithic SRP during verification of a property $P$.
Proofs can be found in Appendix~\ref{sec:proofs}.

We start by proving that the solutions of the fragments $T_{1},T_{2}$
are a solution to the monolithic SRP $S$:
each node of $S$ is mapped to its fragment solution,
with $S$'s input nodes mapping to their expected assumptions.

\begin{theorem}[$\cut$ is Sound]\label{thm:soundness}
  Let $S$ be an open SRP, and let $I$ be an interface over $S$.
  Let $\cut(S, I) = (T_{1}, T_{2})$.
  Suppose $T_{1}$ has a unique solution $\compof{1}{\lab}$ and $T_{2}$ has a unique solution $\compof{2}{\lab}$.
  Consider a mapping $\compof{S}{\lab}' : \compof{S}{V} \rightarrow R$, defined such that:
  \begin{align*}
    \forall v \in \compof{1}{V}.~\compof{S}{\lab}'(v) &= \compof{1}{\lab}(v) \\
    \forall v \in \compof{2}{V}.~\compof{S}{\lab}'(v) &= \compof{2}{\lab}(v) \\
    \forall v \in \compof{S}{\vin}.~\compof{S}{\lab}'(v) &= \compof{S}{\inh}(v)
  \end{align*}
  Then $\compof{S}{\lab}'$ is a solution of $S$.
\end{theorem}

We can also always find a suitable interface $I$ to cut $S$,
such that $T_{1}$ and $T_{2}$ have the same solution as $S$ for each node:
we simply annotate each cut edge $uv$ with the solution $\compof{S}{\lab}(u)$,
which would be the solution transferred from $u$ to $v$ in $S$.

\begin{theorem}[$\cut$ is Complete]\label{thm:completeness}
  Let $S$ be an open SRP, and let $I$ be an interface over $S$.
  Let $\cut(S, I) = (T_{1}, T_{2})$.
  Assume $S$ has a unique solution $\compof{S}{\lab}$.
  Assume that $\forall uv \in \dom(I).~I(uv) = \compof{S}{\lab}(u)$.
  Consider the following two mappings $\compof{1}{\lab}' : \compof{1}{V} \rightarrow R$
  and $\compof{2}{\lab}' : \compof{2}{V} \rightarrow R$, defined such that:
  \begin{align*}
    \forall v \in \compof{1}{V}.~\compof{1}{\lab}'(v) &= \compof{S}{\lab}(v) \\
    \forall v \in \compof{2}{V}.~\compof{2}{\lab}'(v) &= \compof{S}{\lab}(v)
  \end{align*}
  Then $\compof{1}{\lab}'$ is a solution for $T_{1}$ and
  $\compof{2}{\lab}'$ is a solution for $T_{2}$.
\end{theorem}

Finally, our proof of soundness implies that any property that holds over the solutions of
our fragments will hold over the solutions of our monolithic network.

\begin{corollary}[$\cut$ Preserves Properties]\label{cor:properties}
  Let $S$ be an open SRP, and let $I$ be an interface over $S$.
  Let $\cut(S, I) = (T_{1}, T_{2})$.
  Let $P_{1}, P_{2}$ be formulas such that
  $P_{1} = \forall v \in \compof{1}{V}.~Q(v)$ and
  $P_{2} = \forall v \in \compof{2}{V}.~Q(v)$, where $Q$ is a predicate on $\lab(v)$.
  Assume $S$ has a unique solution $\compof{S}{\lab}$,
  and that $T_{1}$ has a solution $\compof{1}{\lab}$
  and $T_{2}$ has a solution $\compof{S}{\lab}$.
  Then if $P_{1}$ holds on $T_{1}$ and $P_{2}$ holds on $T_{2}$,
  $P_{1} \wedge P_{2}$ holds on $S$.
\end{corollary}

\section{Checking Fragments in SMT}\label{sec:algorithm}

We now present our three-step modular verification methodology:
\begin{enumerate*}[label=\emph{(\roman*)}]
	\item given an SRP $S$ and an interface $I$, produce $N$ fragments using $\cut(S, I)$,
	as defined in \S\ref{sec:interfaces}; then
	\item encode each fragment to SMT and check its guarantees and a safety property $P$ under the given assumptions; and
	\item if any guarantees fail, let the user \emph{refine} $I$ or correct network bugs.
\end{enumerate*}
By our theoretical results,
when our SMT solver verifies $P$ for these smaller fragments,
we can conclude that it would have verified $P$ for the monolithic SRP.

\para{Creating Interfaces}\label{sec:creating-interfaces}
For now, we treat our interfaces as given, meaning they
function similarly to user-provided annotations in an annotation
checking tool such as Dafny~\cite{leino2010dafny}.
Hence, our checking algorithm acts as an analogous tool to
verify beliefs about the network.
In this sense, interfaces are user-provided specifications
to the verifier.

Another way to create interfaces is to \emph{infer} them:
starting from a small amount of given information,
say the initial route to a single destination, we
could infer routes through the rest of the network.
While we do not yet consider interface inference,
we believe it is a fruitful direction for future work,
and discuss doing so in~\S\ref{sec:future-work}.

\algrenewcommand\algorithmicfunction{\textbf{proc}}
\begin{algorithm}[t]
	\caption{\label{alg:checker} The fragment checking algorithm.}
	\begin{algorithmic}[1]
		\Function{Solve}{fragment $T$, property $P$}\label{alg:solve}
		\State $N \gets \encode(T)$
		\Comment{closed SRP $\compof{T}{\lab}$ constraints~\eqref{eq:srp-sol}}\label{alg:encode}
		\State $A \gets \bigwedge_{u \in \compof{T}{\vin}} \compof{T}{\lab}(u) = \compof{T}{\inh}(u)$
		\Comment{$\inh$ constraints~\eqref{eq:open-sol-in}}\label{alg:assume}
		\State $G \gets \bigwedge_{u \in \compof{T}{\vout}}\compof{T}{\lab}(u) = \compof{T}{\outh}(u)$
		\Comment{$\outh$ constraints~\eqref{eq:open-sol-out}}\label{alg:guarantee}
		\State \Return \Call{AskSat}{$A \wedge N \wedge \neg (G \wedge P)$}\label{alg:query}
		\EndFunction
		\Statex
		\Function{Check}{SRP $S$, property $P$, interface $I$}\label{alg:check}
		\State $T_{1}, \ldots, T_{N} \gets \cut(S, I)$
		\For{$i \gets 1,N$}
		\State $r \gets $\Call{Solve}{$T_{i}, P$}
		\If{$r \neq \smtunsat$}
			\State \Return $r$
		\EndIf
		\EndFor
		\State \Return \smtunsat
		\EndFunction
	\end{algorithmic}
\end{algorithm}

\para{The Fragment Checking Algorithm}\label{sec:checking-alg}
Algorithm~\ref{alg:checker} shows how we cut an SRP
and check the three constraints on open SRP solutions (described in \S\ref{sec:open-sol})
on each of the fragments.
We start in the \smtcheck procedure on line \algref{alg:checker}{alg:check}.
\smtcheck{} calls $\cut(S, I)$ to cut $S$ into fragments, and then
calls \smtsolve{} (line \algref{alg:checker}{alg:solve}) on each fragment,
reporting any \smtsat result it receives back from the solver.
\smtsolve{} encodes~\eqref{eq:open-sol} on line \algref{alg:checker}{alg:encode},
\eqref{eq:open-sol-in} on line \algref{alg:checker}{alg:assume} and
\eqref{eq:open-sol-out} on line \algref{alg:checker}{alg:guarantee}.
Since we are interested in knowing if $G$ or $P$ are ever violated,
our final formula is the conjunction of $\encode(T)$
and $A$ with the negation of $G \wedge P$ (line \algref{alg:checker}{alg:query}).
\textsc{AskSat} asks our solver if this formula is satisfiable,
and returns either \smtsat with a model, or \smtunsat.
This model will be a quasi-solution $\compof{T}{\lab}$ to $T$
where the $\encode(T)$ and $A$ constraints hold, but
$\exists u \in \compof{T}{\vout}.~\compof{T}{\lab}(u) \not= \compof{T}{\outh}(u)$
(guarantee violation) or $\exists u \in \compof{T}{V}.~\neg P(u)$ (property violation).
Otherwise, if the solver returns \smtunsat,
then either $S$ has no solution or the guarantees and property always hold.

\para{Refining Interfaces}\label{sec:refining-interfaces}
If every fragment returns \smtunsat, by Corollary~\ref{cor:properties},
we conclude that if there exists a solution to each fragment, then
$P$ and $G$ hold and the interface is \emph{correct}.
On the other hand, if any fragment returns \smtsat, we must determine why our property
or guarantees were violated.
For example, in \S\ref{sec:overview},
we considered if our interface correctly captured the intended
network behaviour, but a bug in the network policy led to a guarantee violation.
If the reverse were true --- the network was configured correctly, but our interface is incorrect ---
we must \emph{refine} our interface to correct it.

By Theorem~\ref{thm:soundness}, we know that any incorrect interface will not define a solution
in $T_{1}$ and $T_{2}$, meaning our guarantee constraint in \smtsolve{} fails and a
counterexample is returned.
This counterexample may then inform a new interface we can provide in a successive run of
\smtcheck{}.
Returning to our fattree fragments in Figure~\ref{fig:fat20frags},
suppose our interface provided the incorrect annotation $I(a_{0}c_{0}) = 1$.
This generates an unattainable guarantee $\compof{0}{\outh}(a_{0}) = 1$,
meaning we can reach $d$ in one hop from $a_{0}$.
$\smtsolve(T_{p0}, P)$ returns \smtsat, providing
$\compof{0}{\lab}(a_{0}) = 3$ as a counterexample which violates this guarantee.
We can then create a new interface with $I(a_{0}c_{0}) = 3$ and re-run verification:
if no further annotations are incorrect, then $\smtsolve(T_{p0}, P)$ will report \smtunsat.

\section{Implementation}\label{sec:implementation}

Our \sysname{} extension adds \texttt{partition} and \texttt{interface} functions
to the \nv{} language: when a user runs \nv{} on a file that declares these functions,
\nv{} cuts the SRP into a set of fragments as part of
a \emph{partitioning} step.
Each fragment is generated as described by Definition~\ref{defn:cut} of $\cut$.
Most of the partitioning step deals with restricting the monolithic $\init$ and $\transfer$ functions.
We create $N$ copies of the initial \nv{} file and traverse the AST of each to update any
references to the topology.
This implementation is currently not optimized and performs redundant work, which can be improved
to reduce overhead when partitioning large policies.

Beyond assigning nodes and edges to fragments using \texttt{partition} and \texttt{interface},
\sysname{} also decomposes the properties we wish to test.
Many useful end-to-end properties can be expressed as predicates over individual node solutions,
including reachability, path length, waypointing, black holes and fault tolerance~\cite{minesweeper}.
These properties can be decomposed into separate assertions over the nodes of each fragment:
if no fragment reports a property violation, we can then conclude that the property holds for
the monolithic network as well, as proven in Corollary~\ref{cor:properties}.

\sysname{}'s SMT encoding follows Algorithm~\ref{alg:checker}, using \nv{}'s monolithic SRP encoding
as the encoding function $\encode$.
As discussed above, the solver returns a \smtsat or \smtunsat response for each fragment to the user.
Any fragments that return \smtsat provide a solution violating the guarantees or properties,
allowing us to determine if the violation indicates a problem with our network policy or interface.

\section{Evaluation}\label{sec:evaluation}

We evaluated \sysname{} on a variety of \nv{} benchmarks representing fattree, random and
Internet topologies.%
\footnote{All of our benchmarks are available or adapted from those at~\cite{nv:github}.}
Our questions focus on the scalability and performance of \sysname{} in comparison to \nv{},
specifically:
\begin{enumerate*}[label=\emph{(\roman*)}]
  \item does \sysname{} improve on \nv{} verification time across topologies and properties, and
  \item how do different cuts impact \sysname{} performance?
\end{enumerate*}
We consider two metrics for verification time:
the maximum time reported to verify
an SMT query encoding the monolithic network or fragment using the \zthree{}~\cite{z3} SMT solver;%
\footnote{We take the maximum query time as each fragment SMT query is independent of the others
  and hence could be parallelized on a multi-processor platform or a cluster of servers.}
and the ``total time'' of \nv{},
which is the time taken by \nv{}'s pipeline of network transformations,
partitioning (for cut networks), encoding to SMT and solving every query sequentially.

We ran each benchmark on a computing cluster node with a 2.4GHz processor and up to 24GB of memory per benchmark.
Each benchmark tested verification of either the monolithic network or a cut network,
and we ran each benchmark for 5 trials and took the average time.
We used two timeouts:
an 8-hour timeout on \nv{} as a whole and a 2-hour timeout on \zthree{}.
The \nv{} timeout prevented fragments from spending too long partitioning or solving multiple \zthree{}
queries, while the \zthree{} timeout also ensured that benchmarks did not spend too long solving any single
fragment's SMT query.

\para{Fattrees}\label{sec:fattrees}
To evaluate \sysname{}'s performance for fattrees,
we made use of the shortest path policy \textsf{SP} and valley-free policy \textsf{FAT} described in~\cite{giannarakis2020nv},
along with an original fault-tolerance policy \textsf{MAINT}.
\textsf{MAINT} extends \textsf{SP}
by requiring that nodes avoid routing through a non-destination node \texttt{down}
which is currently down for maintenance:
routes advertised by \textsf{down} will be dropped.
We encode \textsf{down} as a symbolic value, meaning
that we check that routing bypasses the down node \emph{for all concrete choices} of \texttt{down}.

As in~\cite{giannarakis2020nv}, we parameterize fattree designs by $k$, the number of pods: we vary the topology size from
$k=4$ (20 nodes) to $k=20$ (500 nodes) to assess scalability.
Furthermore, we consider four different cuts of our fattree networks:
\begin{itemize}
  \item \textbf{Vertical}: creates 2 fragments, each with half the spines and half the pods;
  \item \textbf{Horizontal}: creates 3 fragments: the pod containing the routing destination,
        the spines, and all the other pods;
  \item \textbf{Pods}: creates $k+1$ fragments (given $k$ pods): the spines and each pod in their own fragment; and
  \item \textbf{Full}: creates $\abs{V}$ fragments (given $\abs{V}$ nodes), with every node in its own fragment.
\end{itemize}
We generate interfaces automatically using shortest paths algorithms, irrespective of the kind of cut.
For \textsf{SP}, a standard shortest-paths computation is sufficient;
for \textsf{FAT}, we track the level of a route to block valleys~\cite{gao2001inferring,valleyfreedc};
and for \textsf{MAINT}, we use Yen's 2-shortest paths algorithm~\cite{yen}:
this gives us the shortest and second-shortest path (taken if \texttt{down} lies on the shortest path)
to the destination from each node.
For our interface, we then assign a route to each cut edge depending on
the value of \texttt{down}.

\pgfplotstableread[search path={./data/,../data/}]{sp-z3-01-14.dat}\sptbl
\pgfplotstableread[search path={./data/,../data/}]{sp-part-09-29.dat}\spparttbl
\pgfplotstableread[search path={./data/,../data/}]{sp-e2e-01-14.dat}\spendtbl
\pgfplotstableread[search path={./data/,../data/}]{fat-merged-z3-01-18.dat}\fattbl
\pgfplotstableread[search path={./data/,../data/}]{fat-merged-e2e-01-18.dat}\fatendtbl
\pgfplotstableread[search path={./data/,../data/}]{maint-z3-01-14.dat}\mainttbl
\pgfplotstableread[search path={./data/,../data/}]{maint-e2e-01-14.dat}\maintendtbl
\captionsetup[sub]{font=normal} 
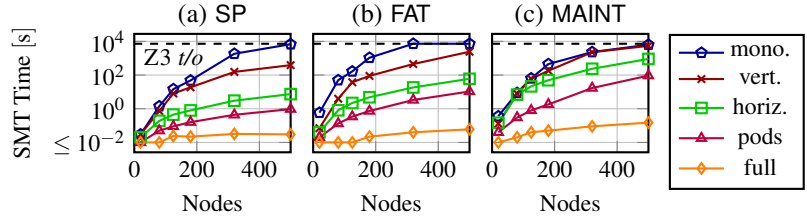
\begin{figure*}[t]
  \centering
  \begin{tikzpicture}
    \begin{groupplot}[
      group style={
        group size=3 by 1,
        xlabels at=edge bottom,
        ylabels at=edge left,
        yticklabels at=edge left,
        horizontal sep=3mm,
      },
        width=0.3\linewidth,
        grid=major,
        xmin=0,
        xmax=500,
        ymode=log,
        ytickten={-2,0,2,4},
        yticklabels={$\leq 10^{-2}$, $10^{0}$, $10^{2}$, $10^{4}$},
        xlabel=Nodes,
        ylabel=SMT Time,
        y unit={\si{\second}},
        cycle list={
          {blue,mark=pentagon},
          {red,mark=x},
          {green,mark=square},
          {purple,mark=triangle},
          {orange,mark=diamond},
        },
        title style={
          text width=0.3\linewidth,
        },
      ]
    \nextgroupplot[thick, title={\subcaption{\label{fig:sp-solve-time}\textsf{SP}}}]
      \foreach \cut in {monolithic,vertical,horizontal,pods,full} {
        \addplot table[x=nodes,y=\cut] {\sptbl};
      }
      \addplot[mark=none,black,dashed,samples=2,domain=0:500] {7200.0}
        node [pos=0.25,yshift=-5pt,font=\it\footnotesize] {\zthree{} t/o};
    \nextgroupplot[thick,title={\subcaption{\label{fig:fat-solve-time}\textsf{FAT}}}]
      \foreach \cut in {monolithic,vertical,horizontal,pods,full} {
        \addplot table[x=nodes,y=\cut] {\fattbl};
      }
      \addplot[mark=none,black,dashed,samples=2,domain=0:500] {7200.0};
    \nextgroupplot[thick,title={\subcaption{\label{fig:maintenance-solve-time}\textsf{MAINT}}},legend style={at={(2.0,1.0)}}]
      \addplot table[x=nodes,y=monolithic] {\mainttbl};
      \addlegendentry{mono.}
      \addplot table[x=nodes,y=vertical] {\mainttbl};
      \addlegendentry{vert.}
      \addplot table[x=nodes,y=horizontal] {\mainttbl};
      \addlegendentry{horiz.}
      \addplot table[x=nodes,y=pods] {\mainttbl};
      \addlegendentry{pods}
      \addplot table[x=nodes,y=full] {\mainttbl};
      \addlegendentry{full}
        \addplot[mark=none,black,dashed,samples=2,domain=0:500] {7200.0};
  \end{groupplot}
  \end{tikzpicture}
  \caption{SMT solve times for fattree benchmarks.}\label{fig:smt-fat}
\end{figure*}

\pgfplotstableread[search path={./data/,../data/}]{fat20-e2e-cmp.dat}\cmpendtbl
\pgfplotstableread[search path={./data/,../data/}]{fat20-z3sum-cmp.dat}\cmpsmttbl
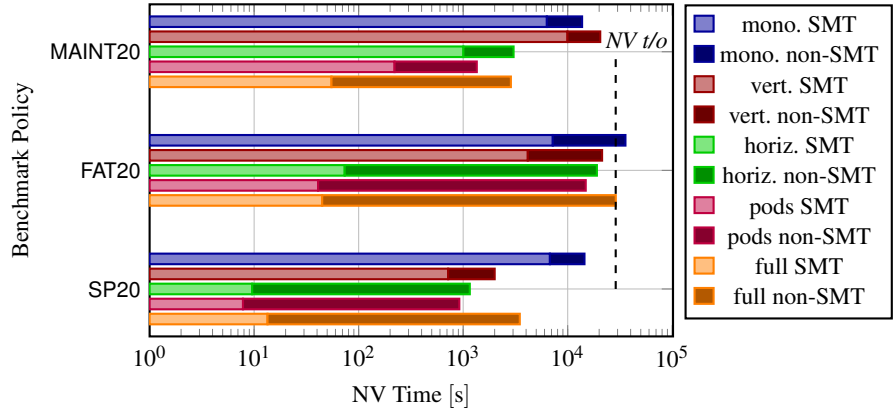
\begin{figure*}[t]
  \centering
  \begin{tikzpicture}
    \begin{axis}[
      grid=major,
      style={thick},
      xbar stacked,
      reverse stacked plots=false,
      bar width=4pt,
      enlarge y limits=0.2,
      ylabel=Benchmark Policy,
      symbolic y coords={SP, FAT, MAINT},
      ytick=data,
      yticklabels={\textsf{SP20}, \textsf{FAT20}, \textsf{MAINT20}},
      xlabel=\nv{} Time,
      xmode=log,
      x unit={\si{\second}},
      xmin={1e0},
      nodes near coords align={horizontal},
      width=0.7\linewidth,
      height=6cm,
      cycle list={
          {blue,fill=blue!50},
          {blue,fill=blue!70!black},
          {red,fill=red!50},
          {red,fill=red!70!black},
          {green,fill=green!50},
          {green,fill=green!70!black},
          {purple,fill=purple!50},
          {purple,fill=purple!70!black},
          {orange,fill=orange!50},
          {orange,fill=orange!70!black},
        },
      legend style={at={(1.42,1.0)}}
      ]
        \addplot +[bar shift=.4cm] table[y=benchmark,x=monolithic] {\cmpsmttbl};
        \addlegendentry{mono. SMT}
        \addplot +[bar shift=.4cm] table[y=benchmark,x=monolithic] {\cmpendtbl};
        \addlegendentry{mono. non-SMT}
        \resetstackedplots

        \addplot +[bar shift=.2cm] table[y=benchmark,x=vertical] {\cmpsmttbl};
        \addlegendentry{vert. SMT}
        \addplot +[bar shift=.2cm] table[y=benchmark,x=vertical] {\cmpendtbl};
        \addlegendentry{vert. non-SMT}
        \resetstackedplots

        \addplot table[y=benchmark,x=horizontal] {\cmpsmttbl};
        \addlegendentry{horiz. SMT}
        \addplot table[y=benchmark,x=horizontal] {\cmpendtbl};
        \addlegendentry{horiz. non-SMT}
        \resetstackedplots

        \addplot +[bar shift=-.2cm] table[y=benchmark,x=pods] {\cmpsmttbl};
        \addlegendentry{pods SMT}
        \addplot +[bar shift=-.2cm] table[y=benchmark,x=pods] {\cmpendtbl};
        \addlegendentry{pods non-SMT}
        \resetstackedplots

        \addplot +[bar shift=-.4cm] table[y=benchmark,x=full] {\cmpsmttbl};
        \addlegendentry{full SMT}
        \addplot +[bar shift=-.4cm] table[y=benchmark,x=full] {\cmpendtbl};
        \addlegendentry{full non-SMT}
        \resetstackedplots

        \addplot[mark=none,black,dashed,const plot] coordinates {
          (28800.0,[normalized]0)
          (28800.0,[normalized]2)
        }
          node [pos=1.05,xshift=8pt,font=\it\footnotesize] {\nv{} t/o};
    \end{axis}
  \end{tikzpicture}
  \caption{\nv{} total times for 20-pod fattree benchmarks.}
  \label{fig:e2e-fat}
\end{figure*}

We compare the SMT verification time for monolithic benchmarks versus their cut counterparts in
Figure~\ref{fig:smt-fat} for each of our policies.
We plot the number of nodes in the monolithic benchmark against the maximum time spent by \zthree{} solving the SMT queries:
for monolithic networks, there is only a single query, while for cut networks, we have $N$ queries.
Note that SMT time is shown on a logarithmic scale.
All three policies show extreme improvements in SMT time as the number of fragments grows.
The maximum SMT time for a full cut fragment of our largest \textsf{SP} network considered
is \emph{six orders of magnitude} faster than the baseline monolithic time.
The \textsf{FAT} policy's SMT encoding is most complex, leading to \zthree{} timeouts for
the monolithic \textsf{FAT16} and \textsf{FAT20} benchmarks.

We compare total time (\IE with no parallelization) in Figure~\ref{fig:e2e-fat}
for the largest fattree benchmarks of each of our three policies for different cuts.
The relationship between cuts is similar for the smaller benchmarks.
We distinguish SMT time from non-SMT (partitioning, encoding, \ETC) time, and see
that across policies, SMT time takes up the majority of total time for the monolithic
benchmarks and vertically-cut benchmarks, but other operations then dominate for the
remaining cuts.
This leads the horizontal and pods cuts to perform the best overall relative to the monolithic
benchmark, completing 2--8x faster.
The \textsf{FAT20} full cut benchmark times out partway through solving due to the time spent
partitioning and encoding, but all other cuts complete before the \nv{} timeout.
As mentioned above, \nv{}'s partitioning step is under-optimized:
hence, we consider slowness outside SMT to be surmountable following improvements
to \sysname{}'s partitioning and \nv{}'s encoding steps.

\para{Random Networks}\label{sec:rand-networks}
We also assess \sysname{} on random networks.
We generate topologies of $N$ nodes using the Erdős–Rényi–Gilbert model~\cite{erdos1959random,gilbert1959random},
where each edge has independent probability $p$ of being present.
To assess scalability, we vary $N$ and $p$ in our experiments according to a parameter $x$ where $N=2^{x}$ and $p=2^{2-x}$
for $x \in [4,12]$%
\footnote{As our topologies are not always fully connected, we expect \nv{} to return property violations as appropriate,
and otherwise for all checks to pass.}.
We use a shortest-path policy based on \textsf{SP} for these networks.
Our interfaces are generated by a shortest paths algorithm and cut the network fully.

\pgfplotstableread[search path={./data/,../data/}]{rand-z3-01-19.dat}\randsmttbl
\begin{figure}[t]
  \centering
  \begin{tikzpicture}
    \begin{axis}[
      width=0.7\linewidth,
      height=3cm,
      grid=major,
      style={thick},
      xlabel=Nodes,
      ylabel=SMT Time,
      ymode=log,
      xmode=log,
      log basis x=2,
      y unit={\si{\second}},
      ytickten={-2,0,2,4},
      yticklabels={$\leq 10^{-2}$, $10^{0}$, $10^{2}$, $10^{4}$},
      cycle list={
          {blue,mark=pentagon},
          {orange,mark=diamond},
        },
      legend style={at={(1.3,1.0)}}
      ]
        \addplot table[x=nodes,y=monolithic] {\randsmttbl};
        \addlegendentry{mono.}
        \addplot table[x=nodes,y=full] {\randsmttbl};
        \addlegendentry{full}
      \addplot[mark=none,black,dashed,samples=2,domain=16:4096] {7200.0}
        node [pos=0.25,yshift=-5pt,font=\it\footnotesize] {\zthree{} t/o};
    \end{axis}
  \end{tikzpicture}
  \caption{Largest SMT solve times for random networks.}
  \label{fig:smt-rand}
\end{figure}
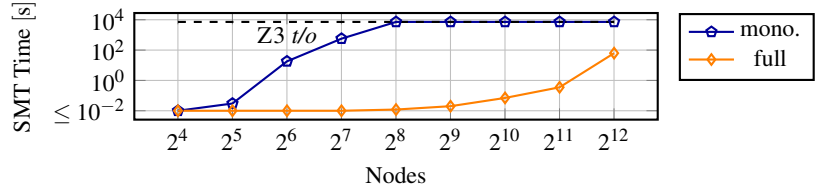

We show the SMT solve times for these benchmarks in Figure~\ref{fig:smt-rand}.
As expected, monolithic verification hits our \zthree{} timeout at $N=256$; fully partitioning
allows us to verify all larger benchmarks in under 6 minutes over \emph{all} SMT queries, with no
individual query taking longer than a minute.

\para{Backbone Networks}\label{sec:backbone-networks}
To assess \sysname{} more fully, we expand our evaluation to backbone network topologies
from the Internet Topology Zoo~\cite{topologyzoo}.
We consider three networks:
a 41-node topology \textsf{B41},
a 174-node topology \textsf{B174} and 754-node topology \textsf{B754}.
\textsf{B41} is an educational network with a more clustered topology:
its policy uses shortest-path routing where routes transiting~\cite{gao2001inferring} through AS customers or peers is disallowed.
The larger topologies are less structured and hence use standard shortest-path routing as in \textsf{SP}.
We use a graph partitioning tool, hMETIS~\cite{hmetis}, to compute $N$ fragments of each topology.
The computed fragments minimize the number of edges cut between fragments, and capture
clustering behavior of the topology, while keeping fragments as close in graph order as possible.
We consider $N = 2, 4, 7, 41$ for \textsf{B41}, $N = 2, 4, 20, 174$ for \textsf{B174}, and $N = 2, 4, 8, 25, 75, 754$
for \textsf{B754}.

\pgfplotstableread[search path={./data/,../data/}]{topzoo-corrected-z3-01-18.dat}\topzoosmttbl
\pgfplotstabletranspose[colnames from=nodes,input colnames to=cut]\topzoosmttrans\topzoosmttbl
\pgfplotstablesort[sort key=cut]\topzoosmtfinal\topzoosmttrans
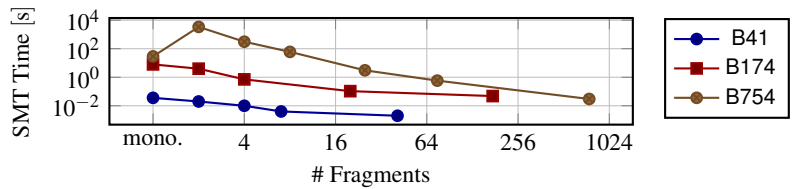
\begin{figure}[t]
  \centering
  \begin{tikzpicture}
    \begin{axis}[
        width=0.7\linewidth,
        height=3cm,
        grid=both,
        style={thick},
        ylabel=SMT Time,
        y unit=\si{\second},
        ymode=log,
        ytickten={-2,0,2,4},
        xmode=log,
        log basis x=2,
        xticklabels={mono., 4, 16, 64, 256, 1024},
        xlabel=\# Fragments,
        legend style={at={(1.3,1.0)}},
      ]
      \addplot table[x=cut,y=41] {\topzoosmtfinal};
      \addlegendentry{\textsf{B41}};
      \addplot table[x=cut,y=174] {\topzoosmtfinal};
      \addlegendentry{\textsf{B174}};
      \addplot table[x=cut,y=754] {\topzoosmtfinal};
      \addlegendentry{\textsf{B754}};
    \end{axis}
  \end{tikzpicture}
  \caption{\label{fig:topologyzoo-smt} Largest SMT solve times for TopologyZoo networks.}
\end{figure}

We show that larger cuts lead to greater reductions in SMT solve time for these benchmarks in Figure~\ref{fig:topologyzoo-smt}.
As for fattrees, the lowest total times tend to be lowest for
larger non-full cuts ($N=20$ for \textsf{B174} and $N=75$ for \textsf{B754}).

\section{Related Work}\label{sec:related}

\para{Data Plane Analysis}
Much prior work has analyzed properties of the network data
plane~\cite{anteater,hsa,veriflow,netplumber,netkat,nod,surgery,secguru}.
These tools operate on snapshots of the data plane --- representing the global forwarding
state at a single point in time --- and verify that forwarding properties are satisfied.

Our approach most closely resembles the work of Jayaraman \ETAL{} on \secguru{} and RCDC~\cite{secguru}.
\secguru{} verifies reachability using invariants it infers from specific data center topologies:
our work develops a formal theory to verify arbitrary properties and invariants as specified by a user's interface,
provides a framework for doing so automatically
and instead focuses on the control plane.

Another relevant work is that of Plotkin \ETAL~\cite{surgery}.
They demonstrate the use of bisimulations to relate simpler networks and formulas
to more complex ones, improving verification scalability.
Modular verification is recognized as a viable direction but left as future work; we
focus on using modular verification in the control plane.

\para{Control Plane Analysis}
Our open SRP model builds on prior work on formal models of control planes:
in particular, the SRP model of Bonsai~\cite{bonsai},
which presents a topology with an attached routing algebra.
Unlike other prior work~\cite{daggitt2018asynchronous,stable-paths,metarouting},
we ignore questions of network convergence and assume a unique solution exists to our network.

Many control plane verification tools address scalability
by abstracting routing behaviors, rather than modularizing the network.
Abstraction necessarily loses precision,
which can limit the properties or networks considered.
Bonsai~\cite{bonsai} and Origami~\cite{origami} are perhaps closest to our work
in that they seek to \emph{compress} large concrete networks to smaller abstract networks
which soundly approximate the original.
Both tools use abstraction refinement to find abstract networks and use a similar
formal model to our own.
Compression requires similar forwarding behavior across multiple nodes of the network;
our cutting approach avoids this restriction.

ShapeShifter~\cite{shapeshifter} checks control plane reachability
by simulating the network using abstract routes determined by abstract interpretation.
They define an asynchronous network semantics for routing; we instead model the
network's converged state using open SRPs.
ShapeShifter's abstractions sacrifice precision, unlike our technique.

Our SMT encoding is inspired by Minesweeper~\cite{minesweeper}, although
we do not consider packet forwarding (only routing) and Minesweeper cannot perform modular verification.
Plankton~\cite{plankton} uses explicit-state model checking to check a comparable set of properties
to Minesweeper and use a network semantics similar to ShapeShifter's.
They avoid state explosion using heuristic reductions that work well for the networks considered.
Our approach is more general and avoids explicitly exploring network states by using SMT.
Other control plane analyses also do not consider modularizing the network,
and many are more restrictive than our approach:
either limited to specific network properties~\cite{arc,tiramisu,era} or
to specific protocols~\cite{bagpipe}.

\para{Modular Verification}
As mentioned above, our work borrows from the compositional verification technique of
\emph{assume-guarantee reasoning}~\cite{alur1999reactive,flanagan2003thread,giannakopoulou2018compositional}.
Such reasoning has been widely used in software, hardware and reactive
systems~\cite{henzinger1998you,flanagan2003thread,grumberg1994model}.
While~\cite{lomuscio2010assume} applies assume-guarantee reasoning in
network congestion control, assume-guarantee
appears to be under-explored in analyzing routing.
Instead of modeling processes, we model network fragments,
whose shared environment is their input and output nodes.
By requiring a partition's assumptions and guarantees to be equal,
our reasoning avoids the common pitfall of circularity by
relying on the stability of an open SRP's solution.

\section{Discussion \& Future Work}\label{sec:discussion}

\para{Choosing Cuts}\label{sec:choosing-cuts}
This paper answers two major questions about network partitions: first, given a network cut,
can we verify properties of a monolithic network using its fragments?
Second, does verifying the fragments scale better than verifying the monolithic network?
We leave unanswered a third critical question: \emph{where} should we cut?

As we saw in our evaluation, verification time is inversely proportional to the number of fragments.
However, this introduces a tradeoff: for every edge our interface cuts, we must supply another annotation.
How easy it may be to annotate a given edge depends on many factors:
who manages the network (\EG private organizations \VS the internet), how policy is determined along the edge, \ETC
Nonetheless, by making these factors explicit using interfaces, we make it easier to understand the monolithic
behavior of legacy networks, thereby improving their safety and long-term robustness.

\para{Future Directions}\label{sec:future-work}
Our theoretical framework provides a foundation for
two promising avenues for future work: \emph{abstracting} our interface annotations
and \emph{inferring} interfaces automatically.
Using abstract annotations
--- annotating with a set of routes, \EG $I(uv) = \{n~\vert~0 \leq n \leq 4\}$, instead of a concrete route $I(uv) = 4$ ---
we could potentially simplify the task of annotating the network by
over-approximating the set of routes we assume.
Interface inference, perhaps building on our manual refinement process
in \S\ref{sec:refining-interfaces}, could further simplify this task:
we hypothesize a viable technique might use counterexamples to automatically
refine our initial interface as a series of verification passes~\cite{ClarkeCegar00}.

\para{Conclusion}\label{sec:conclusion}
We demonstrate that scalability in control plane verification can be achieved
by leveraging networks' inherent modularity.
We prove that we can verify a property of a network by
verifying it independently across fragments of the original,
and present a procedure to do so.
We implemented this procedure in \nv{} as \sysname{} and show that it succeeds
in verifying \nv{} benchmarks with dramatic improvements in SMT time.

\para{Acknowledgements}
This work was supported in part by the National Science Foundation awards NeTS 1704336 and FMitF 1837030,
and Facebook Research Award on “Network control plane verification at scale.”
We would like to thank Todd Millstein for his contributions in our early discussions.
Our evaluation was substantially performed using the Princeton Research Computing
resources at Princeton University, which is a consortium of groups led by the
Princeton Institute for Computational Science and Engineering
(PICSciE) and the Office of Information Technology's Research Computing.
%
%
%
\bibliographystyle{splncs04}
\bibliography{references}
\clearpage
\appendix

\section{Proofs}\label{sec:proofs}

\para{Cuts Form Partitions}
We start by stating the formal definition of a cut from \S\ref{sec:interfaces}.
For our formal definition of $\cut$, we add an additional structural
restriction over our interface $I$ to simplify some of our definitions.
Essentially, we will require that, given an SRP $S$, $I$ cuts $S$
along its base and output nodes.
Formally, for an SRP $S$, let the \emph{input-free graph} of $S$ be
$(\compof{S}{V} \setminus \compof{S}{\vin}, \{ uv ~\vert~ u,v \in \compof{S}{V} \setminus \compof{S}{\vin}\})$,
\IE the induced subgraph of $S$'s base and output nodes.
If we cut the input-free graph into $(W_{1},W_{2})$, we then can assign the input nodes of $S$ to the two fragments
in order to cover $S$:
$\compof{S}{\vin}$ is disjoint from $W_{1}$ and $W_{2}$ and $\compof{S}{V} = W_{1} \cup W_{2} \cup \compof{S}{\vin}$.
Any input node $u \in \compof{S}{\vin}$ which has an edge $uv$ to a node $v$
in $W_{1}$ (respectively $W_{2}$) is also an input node $u \in \compof{1}{\vin}$ (respectively $\compof{2}{\vin}$).
Importantly, if there exists $u, v_{1}, v_{2} \in \compof{S}{V}$ and $uv_{1}, uv_{2} \in \compof{S}{E}$,
if $v_{1} \in W_{1}$ and $v_{2} \in W_{2}$, then $u$ is a \emph{shared input} in both $T_{1}$ and $T_{2}$,
\IE $u \in \compof{1}{\vin} \cap \compof{2}{\vin}$.

\begin{definition}[$\cut$]\label{defn:cut-apx}
  Let $S$ be an SRP.
  Let $(W_{1}, W_{2})$ be a cut of the input-free graph of $S$
  where $C$ is a cut-set of edges $\{uv \in \compof{S}{E} ~\vert~ (u \in W_{1} \wedge v \in W_{2}) \vee (u \in W_{2} \wedge v \in W_{1})\}$.
  Let $I$ be an interface over $S$ such that $\dom(I)$ is equal to $C$.
  Then $\cut(S, I) = (T_{1}, T_{2})$
  where
  the following properties hold for $i \in \{1, 2\}$:
  \begin{align*}
    \compof{i}{\vin} &= \{ u ~\vert~ u \in \compof{S}{\vin} \wedge \exists uv \in \compof{S}{E}.~v \in W_{i}\}
          \cup \{ u ~\vert~ \exists uv \in \dom(I).~v \in W_{i} \} \\
    \compof{i}{\vout} &= \{ u ~\vert~ u \in W_{i} \wedge u \in \compof{S}{\vout} \}
          \cup \{ u ~\vert~ \exists uv \in \dom(I).~u \in W_{i} \} \\
    \compof{i}{V} &= W_{i} \cup \compof{i}{\vin} \\
    \compof{i}{E} &= \{ uv ~\vert~ u,v \in \compof{i}{V} \wedge uv \in \compof{S}{E} \} \\
    \compof{i}{R} &= \compof{S}{R} \\
    \compof{i}{\init} &= \compof{S}{\init} |_{\compof{i}{V}} \\
    \compof{i}{\merge} &= \compof{S}{\merge} \\
    \compof{i}{\transfer} &= \compof{S}{\transfer} |_{\compof{i}{V}} \\
    \compof{i}{\inh}(u) &=
                       \begin{cases}
                         \compof{S}{\inh}(u) &\textit{if}~u \in \compof{S}{\vin} \\
                         I(uv) &\textit{if}~uv \in \dom(I) \wedge v \in \compof{i}{V}
                       \end{cases} \\
    \compof{i}{\outh}(u) &=
                        \begin{cases}
                          \compof{S}{\outh}(u) &\textit{if}~u \in (\compof{S}{\vout} \setminus \compof{i}{\vin}) \\
                          I(uv) &\textit{if}~uv \in \dom(I) \wedge v \notin \compof{i}{V}
                        \end{cases}
  \end{align*}
\end{definition}

We now state the partition relation that summarizes the properties $\cut(S, I)$ ensures.

\begin{definition}[Partition]\label{defn:partition-apx}
  Let $S$, $T_{1}$ and $T_{2}$ be open SRPs.
  $(T_{1}, T_{2})$ is a \emph{partition} of $S$ when:
  \begin{itemize}
    \item $T_{1}$ and $T_{2}$ are both fragments of $S$
    \item $\compof{1}{V} \cup \compof{2}{V} = \compof{S}{V}$ and $\compof{1}{E} \cup \compof{2}{E} = \compof{S}{E}$
    \item Input-output constraints:
          every input or output that is not inherited from the parent is an input-output node:
          \begin{itemize}
            \item $\compof{1}{\vin} \setminus \compof{S}{\vin} \subseteq \compof{2}{\vout}$
            \item $\compof{2}{\vin} \setminus \compof{S}{\vin} \subseteq \compof{1}{\vout}$
            \item $\compof{1}{\vout} \setminus \compof{S}{\vout} \subseteq \compof{2}{\vin}$
            \item $\compof{2}{\vout} \setminus \compof{S}{\vout} \subseteq \compof{1}{\vin}$
            \item $\forall v \in (\compof{1}{\vin} \setminus \compof{S}{\vin}).~\compof{1}{\inh}(v) = \compof{2}{\outh}(v)$
            \item $\forall v \in (\compof{2}{\vin} \setminus \compof{S}{\vin}).~\compof{2}{\inh}(v) = \compof{1}{\outh}(v)$
          \end{itemize}
    \item Shared input constraint: a node shared by $T_{1}$ and $T_{2}$ is either
          an input into both fragments, or an input-output node:
          $\compof{1}{V} \cap \compof{2}{V} = (\compof{1}{\vin} \cap \compof{2}{\vin})
          \cup (\compof{1}{\vin} \cup \compof{2}{\vin}) \setminus \compof{S}{\vin}$
  \end{itemize}
\end{definition}

The properties of a partition state everything we still need (beyond the properties of open SRPs and fragments)
in order to prove that our $\cut$ procedure is correct.
Our input-output constraints state that the input-output nodes
must agree on their assumptions and guarantees,
and that these nodes make up a subset of the respective input and output nodes
in each sibling fragment (since some input and output nodes may be inherited from $S$).
Our shared input constraint states that,
if a node $u$ appears in both $T_{1}$ and $T_{2}$, then $u$ is either
\begin{enumerate*}[label=\emph{(\roman*)}]
  \item an input-output node; or
  \item a shared input of both $T_{1}$ and $T_{2}$.
\end{enumerate*}

We prove that as defined, $\cut(S, I)$ is a partition of $S$.
This is a straightforward proof from the definition of $\cut$,
using some set identities to prove the properties of a partition.

\begin{theorem}[$\cut$ Creates Partitions]\label{thm:cut-partition-apx}
  Let $S$ be an SRP, and let $I$ be an interface over $S$.
  Let $\cut(S, I) = (T_{1},T_{2})$.
  Then $(T_{1},T_{2})$ is a partition of $S$.
\end{theorem}
\begin{proof}
  Consider the input-free graph of $S$,
  $(\compof{S}{V} \setminus \compof{S}{\vin}, \{ uv ~\vert~ u,v \in \compof{S}{V} \setminus \compof{S}{\vin}\})$,
  such that
  $C = (W_{1}, W_{2})$ cuts the input-free graph
  with $\dom(I)$ as the cut-set of $C$.

  It is trivial to see that based on the definition of $\cut$, $T_{1}$ and $T_{2}$ are both fragments
  of $S$: we hence proceed to prove the remaining properties of the partition relation below.

  \begin{align*}
    \compof{1}{V} \cup \compof{2}{V} &= (W_{1} \cup \compof{1}{\vin}) \cup (W_{2} \cup \compof{2}{\vin})
    & \textit{by definition of } I \\
                                     &= (W_{1} \cup W_{2}) \cup (\compof{1}{\vin} \cup \compof{2}{\vin})
    & \textit{by commutativity, associativity} \\
                                     &= (\compof{S}{V} \setminus \compof{S}{\vin}) \cup (\compof{1}{\vin} \cup \compof{2}{\vin})
                                       & \textit{by definition of }W_{1} \cup W_{2} \\
                                     &= (\compof{S}{V} \cup (\compof{1}{\vin} \cup \compof{2}{\vin})) \setminus (\compof{S}{\vin} \setminus (\compof{1}{\vin} \cup \compof{2}{\vin}))
                                       & \textit{by set identity} \\
                                     &= \compof{S}{V} \setminus (\compof{S}{\vin} \setminus (\compof{1}{\vin} \cup \compof{2}{\vin}))
                                       & \textit{by } \compof{1}{\vin} \cup \compof{2}{\vin} \subseteq \compof{S}{V} \\
                                     &= \compof{S}{V} \setminus \emptyset
                                       & \textit{by } \compof{1}{\vin} \cup \compof{2}{\vin} \supseteq \compof{S}{\vin} \\
                                     &= \compof{S}{V}
    & \textit{by definition of } \setminus
  \end{align*}
  Then $\compof{1}{V} \cup \compof{2}{V} = \compof{S}{V}$.

  \begin{align*}
    \compof{1}{E} \cup \compof{2}{E} &= \{ uv ~\vert~ u,v \in \compof{1}{V} \wedge uv \in \compof{S}{E} \}
                                     \cup \{ uv ~\vert~ u,v \in \compof{2}{V} \wedge uv \in \compof{S}{E} \}
                                     & \textit{by definition of } \compof{1}{E}, \compof{2}{E} \\
                                     &= \{ uv ~\vert~ u,v \in \compof{1}{V} \cup \compof{2}{V} \wedge uv \in \compof{S}{E} \}
                                     & \textit{set identities} \\
                                     &= \{ uv ~\vert~ u,v \in \compof{S}{V} \wedge uv \in \compof{S}{E} \}
                                     &\textit{by } \compof{1}{V} \cup \compof{2}{V} = \compof{S}{V} \\
                                     &= \compof{S}{E}
  \end{align*}
  Then $\compof{1}{E} \cup \compof{2}{E} = \compof{S}{E}$.

  For the input-output constraints, we show one side: the other direction is symmetrical.
  \begin{align*}
    \compof{1}{\vin} \setminus \compof{S}{\vin} &= \{ u ~\vert~ \exists uv \in \dom(I).~v \in W_{1} \}
                                                &\textit{by definition of } \compof{1}{\vin} \\
                                                &= \{ u ~\vert~ \exists uv \in \dom(I).~u \in W_{2} \}
                                                &\textit{since }\dom(I) \textit{ is a cut-set} \\
                                                &\subseteq \compof{2}{\vout}
                                                &\textit{by definition of } \compof{2}{\vout} \\
    \compof{1}{\vout} \setminus \compof{S}{\vout} &= \{ u ~\vert~ \exists uv \in \dom(I).~u \in W_{1} \}
                                                &\textit{by definition of } \compof{1}{\vout} \\
                                                &= \{ u ~\vert~ \exists uv \in \dom(I).~v \in W_{2} \}
                                                &\textit{since }\dom(I) \textit{ is a cut-set} \\
                                                &\subseteq \compof{2}{\vin}
                                                &\textit{by definition of } \compof{2}{\vin}
  \end{align*}
  Then $\compof{1}{\vin} \setminus \compof{S}{\vin} \subseteq \compof{2}{\vout}$ and
  $\compof{1}{\vout} \setminus \compof{S}{\vout} \subseteq \compof{2}{\vin}$.
  The other directional is symmetrical, swapping 1 and 2.
  Then the input-output constraints hold.

  Finally, we can determine that the shared input constraint holds as follows:
  \begin{align*}
    \compof{1}{V} \cap \compof{2}{V} &= (W_{1} \cup \compof{1}{\vin}) \cap (W_{2} \cup \compof{2}{\vin})
                                       &\textit{definition of V} \\
                                     &= (W_{1} \cap W_{2}) \cup (W_{1} \cap \compof{2}{\vin}) \cup (\compof{1}{\vin} \cap W_{2})
                                       \cup (\compof{1}{\vin} \cap \compof{2}{\vin})
                                     &\textit{distributivity} \\
                                     &= \emptyset \cup (W_{1} \cap \compof{2}{\vin}) \cup (\compof{1}{\vin} \cap W_{2})
                                       \cup (\compof{1}{\vin} \cap \compof{2}{\vin})
                                     &\textit{disjointness of Ws} \\
                                     &= (\compof{1}{\vin} \cap \compof{2}{\vin})
                                       \cup (\compof{2}{\vin} \setminus (W_{2} \cup \compof{S}{\vin}))
                                       \cup (\compof{1}{\vin} \setminus (W_{1} \cup \compof{S}{\vin}))
                                     &\textit{commutativity, rewrite Ws} \\
                                     &= (\compof{1}{\vin} \cap \compof{2}{\vin})
                                       \cup (\compof{2}{\vin} \setminus W_{2} \setminus \compof{S}{\vin})
                                       \cup (\compof{1}{\vin} \setminus W_{1} \setminus \compof{S}{\vin})
                                     &\textit{set identity} \\
                                     &= (\compof{1}{\vin} \cap \compof{2}{\vin})
                                       \cup (\compof{2}{\vin} \setminus \compof{S}{\vin})
                                       \cup (\compof{1}{\vin} \setminus \compof{S}{\vin})
                                     &\textit{by }\vin \cap W = \emptyset \\
                                     &= (\compof{1}{\vin} \cap \compof{2}{\vin})
                                       \cup (\compof{2}{\vin} \cup \compof{1}{\vin}) \setminus \compof{S}{\vin}
                                     &\textit{factoring}
  \end{align*}

  Then all the partition relation constraints hold, so $T_{1}, T_{2}$ is a partition of $S$.
\end{proof}

\para{Correctness}
We now continue with a series of lemmas we will use in our proofs of soundness and completeness.
As a reminder to readers, our theorems of soundness and completeness
focus on demonstrating that the solutions of an open SRP's fragments are the
solution of the parent SRP (or vice-versa): we prove these theorems by making use of
case analysis over the cases of an open SRP's solution, as presented
in \S\ref{sec:open-sol}.
Loosely speaking, like the three subsets of an SRP's nodes,
these cases can be divided into
\begin{enumerate*}[label=\emph{(\alph*)}]
  \item base node solutions (\CF{} closed SRP solutions);
  \item input node solutions (equality to $\inh$); and
  \item output node solutions (the closed SRP solution plus equality to $\outh$).
\end{enumerate*}
It is straightforward by the definitions of fragments that, if fragments inherit input and output nodes
from their parent, then their solution will
also be a solution in the parent; the more difficult cases involve using the closed SRP solution,
and reasoning over input-output nodes between the two fragments after a parent edge was cut.

Our three following lemmas help us through these difficult cases
by proving properties of the nodes which are in both fragments of a partition.
Lemma~\ref{lem:1} starts by proving that any node in both fragments of
a partition must be either an input node or an output node.

\begin{lemma}[Shared nodes are either inputs or outputs]\label{lem:1}
  Let $S, T_{1}, T_{2}$ be open SRPs such that $(T_{1}, T_{2})$ is a partition of $S$.
  Then $\compof{1}{V} \cap \compof{2}{V} \subseteq \compof{1}{\vin} \cup \compof{1}{\vout}$ and
  $\compof{1}{V} \cap \compof{2}{V} \subseteq \compof{2}{\vin} \cup \compof{2}{\vout}$.
\end{lemma}
\begin{proof}

  \para{$T_{1}$ case}
  $\compof{1}{V} \cap \compof{2}{V} \subseteq \compof{1}{\vin} \cup \compof{1}{\vout}$
  \begin{align*}
    \compof{1}{V} \cap \compof{2}{V} &= (\compof{1}{\vin} \cap \compof{2}{\vin})
                                       \cup (\compof{1}{\vin} \cup \compof{1}{\vin})
                                       \setminus \compof{S}{\vin}
    &\textit{by shared node division constraint} \\
                                     &= (\compof{1}{\vin} \cap \compof{2}{\vin})
                                       \cup (\compof{1}{\vin} \setminus \compof{S}{\vin})
                                       \cup (\compof{2}{\vin} \setminus \compof{S}{\vin})
    &\textit{distribute } \setminus \textit{ over } \cup \\
                                     &\subseteq (\compof{1}{\vin} \cap \compof{2}{\vin})
                                       \cup \compof{1}{\vin}
                                       \cup (\compof{2}{\vin} \setminus \compof{S}{\vin})
    &\textit{by definition of } \subseteq, \setminus \\
                                     &\subseteq (\compof{1}{\vin} \cap \compof{2}{\vin})
                                       \cup \compof{1}{\vin}
                                       \cup \compof{1}{\vout}
    &\textit{by } \compof{2}{\vin} \setminus \compof{S}{\vin} \subseteq \compof{1}{\vout} \\
                                     &\subseteq \compof{1}{\vin}
                                       \cup \compof{1}{\vin}
                                       \cup \compof{1}{\vout}
    &\textit{by definition of } \subseteq, \cap \\
                                     &\subseteq \compof{1}{\vin}
                                       \cup \compof{1}{\vout}
    &\textit{by } \cup \textit{ idempotence}
  \end{align*}

  \para{$T_{2}$ case}
  Similar to the $T_{1}$ case.
\end{proof}

We next prove an additional lemma about shared input nodes in Lemma~\ref{lem:2}:
if a node is an input to both fragments, then it is also an input of the parent SRP\@.

\begin{lemma}[Shared Inputs are Inherited]\label{lem:2}
  Let $T_{1}, T_{2}, S$ be open SRPs such that $(T_{1}, T_{2})$ is a partition of $S$.
  Then $\compof{1}{\vin} \cap \compof{2}{\vin} \subseteq \compof{S}{\vin}$.
\end{lemma}
\begin{proof}
  \begin{align*}
    &\compof{1}{\vin} \cap \compof{1}{\vout} = \emptyset
    &\textit{by definition of open SRPs} \\
    \Rightarrow &\compof{1}{\vin} \cap (\compof{2}{\vin} \setminus \compof{S}{\vin}) = \emptyset
    &\textit{by input-output constraints} \\
    \Rightarrow &(\compof{1}{\vin} \cap \compof{2}{\vin}) \setminus \compof{S}{\vin} = \emptyset
    &\textit{by } A \cap (B \setminus C) = (A \cap B) \setminus C \\
    \Rightarrow &\compof{1}{\vin} \cap \compof{2}{\vin} \subseteq \compof{S}{\vin}
    &\textit{by } A \setminus B = \emptyset \Rightarrow A \subseteq B
  \end{align*}
\end{proof}

We will use Lemma~\ref{lem:2} in the following proof which now moves on to considering
open SRP solutions directly by proving that, if the two fragments have solutions,
then the solutions are equal for shared nodes.

\begin{lemma}[Shared Nodes have the Same Solutions]\label{lem:3}
  Let $T_{1}, T_{2}, S$ be open SRPs such that $(T_{1}, T_{2})$ is a partition of $S$.
  Assume $T_{1}$ has a solution $\compof{1}{\lab}$ and $T_{2}$ has a solution $\compof{2}{\lab}$.
  Then $\forall v \in (\compof{1}{V} \cap \compof{2}{V}).~\compof{1}{\lab}(v) = \compof{2}{\lab}(v)$.
\end{lemma}
\begin{proof}
  We want to show that
  $\forall v \in (\compof{1}{V} \cap \compof{2}{V}).~\compof{1}{\lab}(v) = \compof{2}{\lab}(v)$.
  Recall the shared node division constraint:
  \[
    \compof{1}{V} \cap \compof{2}{V} = (\compof{1}{\vin} \cap \compof{2}{\vin})
        \cup (\compof{1}{\vin} \cup \compof{2}{\vin}) \setminus \compof{S}{\vin}
  \]
  Then, by substitution, we want to show:
  \[
    \forall v \in ((\compof{1}{\vin} \cap \compof{2}{\vin})
        \cup (\compof{1}{\vin} \cup \compof{2}{\vin}) \setminus \compof{S}{\vin}).~\compof{1}{\lab}(v) = \compof{2}{\lab}(v)
  \]
  which we can split into two separate conjuncts:
  \[
    (\forall v \in (\compof{1}{\vin} \cap \compof{2}{\vin}).~\compof{1}{\lab}(v) = \compof{2}{\lab}(v))
    \wedge
    (\forall v \in ((\compof{1}{\vin} \cup \compof{2}{\vin}) \setminus \compof{S}{\vin}).~\compof{1}{\lab}(v) = \compof{2}{\lab}(v))
  \]

  \para{Case 1: $(\forall v \in (\compof{1}{\vin} \cap \compof{2}{\vin}).~\compof{1}{\lab}(v) = \compof{2}{\lab}(v))$}
  Consider an arbitrary $v$ in $\compof{1}{\vin} \cap \compof{2}{\vin}$.
  By Lemma~\ref{lem:2}, $v \in (\compof{1}{\vin} \cap \compof{2}{\vin}) \rightarrow v \in \compof{S}{\vin}$.
  Then $v \in \compof{1}{\vin} \cap \compof{2}{\vin} \cap \compof{S}{\vin}$.
  Then by the definition of a fragment,
  $\compof{1}{\inh}(v) = \compof{S}{\inh}(v)$ and $\compof{2}{\inh}(v) = \compof{S}{\inh}(v)$.
  By transitivity and the definitions of $\compof{1}{\lab}$ and $\compof{2}{\lab}$, we then have
  $\compof{1}{\lab}(v) = \compof{S}{\inh}(v)$ and $\compof{2}{\lab}(v) = \compof{S}{\inh}(v)$.
  Then, again by transitivity, $\compof{1}{\lab}(v) = \compof{2}{\lab}(v)$.

  \para{Case 2: $(\forall v \in ((\compof{1}{\vin} \cup \compof{2}{\vin}) \setminus \compof{S}{\vin}).~\compof{1}{\lab}(v) = \compof{2}{\lab}(v))$}
  Recall that by the input-output constraints, we have the following:
  \begin{align}
    \compof{1}{\vin} \setminus \compof{S}{\vin} &\subseteq \compof{2}{\vout}\label{eq:l3-wfi0} \\
    \compof{2}{\vin} \setminus \compof{S}{\vin} &\subseteq \compof{1}{\vout}\label{eq:l3-wfi1} \\
    \forall v \in (\compof{1}{\vin} \setminus \compof{S}{\vin}).~\compof{1}{\inh}(v) &= \compof{2}{\outh}(v)\label{eq:l3-wfi4} \\
    \forall v \in (\compof{2}{\vin} \setminus \compof{S}{\vin}).~\compof{2}{\inh}(v) &= \compof{1}{\outh}(v\label{eq:l3-wfi5})
  \end{align}
  We also have the following by the definition of $\lab$:
  \begin{align}
    \forall v \in \compof{1}{\vin}.~\compof{1}{\lab}(v) &= \compof{1}{\inh}(v)\label{eq:l3-lab0} \\
    \forall v \in \compof{2}{\vin}.~\compof{2}{\lab}(v) &= \compof{2}{\inh}(v)\label{eq:l3-lab1} \\
    \forall v \in \compof{1}{\vout}.~\compof{1}{\lab}(v) &= \compof{1}{\outh}(v)\label{eq:l3-lab2} \\
    \forall v \in \compof{2}{\vout}.~\compof{2}{\lab}(v) &= \compof{2}{\outh}(v)\label{eq:l3-lab3}
  \end{align}
  Using the relationships between the sets, we can then substitute the equalities over solutions
  into Equations~\eqref{eq:l3-wfi4} and~\eqref{eq:l3-wfi5} to get the desired statement.

  Since $\compof{1}{\vin} \setminus \compof{S}{\vin} \subseteq \compof{2}{\vout}$ by~\eqref{eq:l3-wfi0},
  and $\compof{1}{\vin} \setminus \compof{S}{\vin} \subseteq \compof{1}{\vin}$ by set identities,
  we can substitute $\compof{2}{\lab}$ for $\compof{2}{\outh}$ (per~\eqref{eq:l3-lab3}) and
  $\compof{1}{\lab}$ for $\compof{1}{\inh}$ (per~\eqref{eq:l3-lab0}) in Equation~\eqref{eq:l3-wfi4} to get a statement
  over solutions:
  \[
    \forall v \in (\compof{1}{\vin} \setminus \compof{S}{\vin}).~\compof{1}{\lab}(v) = \compof{2}{\lab}(v)
  \]

  We can use the same reasoning with Equation~\eqref{eq:l3-wfi1} and Equations~\eqref{eq:l3-lab2} and~\eqref{eq:l3-lab1}
  to get another statement from Equation~\eqref{eq:l3-wfi5}:
  \[
    \forall v \in (\compof{2}{\vin} \setminus \compof{S}{\vin}).~\compof{2}{\lab}(v) = \compof{1}{\lab}(v)
  \]

  We can then rearrange the ground formulas by commutativity and conjoin the two statements to obtain:
  \[
    (\forall v \in (\compof{1}{\vin} \setminus \compof{S}{\vin}).~\compof{1}{\lab}(v) = \compof{2}{\lab}(v)) \wedge
    (\forall v \in (\compof{2}{\vin} \setminus \compof{S}{\vin}).~\compof{1}{\lab}(v) = \compof{2}{\lab}(v))
  \]

  Finally, we can rewrite the conjunction to instead be one formula over
  $(\compof{1}{\vin} \setminus \compof{S}{\vin}) \cup (\compof{2}{\vin} \setminus \compof{S}{\vin})$:
  factoring out the set difference gives us:
  $(\forall v \in ((\compof{1}{\vin} \cup \compof{2}{\vin}) \setminus \compof{S}{\vin}).~\compof{1}{\lab}(v) = \compof{2}{\lab}(v))$,
  which was what was required.
\end{proof}

We now move onto the proof of soundness of $\cut$, which states
that if $\cut(S,I) = (T_{1},T_{2})$,
which have respective solutions $\compof{1}{\lab}$ and $\compof{2}{\lab}$,
then there is a solution to the parent SRP $S$
which is equal to both fragment solutions over all relevant nodes.
We define this solution in the theorem statement, and then prove it satisfies
the solution constraints for any node in $S$, regardless of whether it is
an input, an output or a base node.

\begin{theorem}[$\cut$ is Sound]\label{thm:soundness-apx}
  Let $S$ be an open SRP, and let $I$ be an interface over $S$.
  Let $\cut(S, I) = (T_{1}, T_{2})$.
  Suppose $T_{1}$ has a unique solution $\compof{1}{\lab}$ and $T_{2}$ has a unique solution $\compof{2}{\lab}$.
  Consider a mapping $\compof{S}{\lab}' : \compof{S}{V} \rightarrow R$, defined such that:
  \begin{align}
    \forall v \in \compof{1}{V}.~\compof{S}{\lab}'(v) &= \compof{1}{\lab}(v) \label{eq:lab2prime-0} \\
    \forall v \in \compof{2}{V}.~\compof{S}{\lab}'(v) &= \compof{2}{\lab}(v) \label{eq:lab2prime-1} \\
    \forall v \in \compof{S}{\vin}.~\compof{S}{\lab}'(v) &= \compof{S}{\inh}(v) \label{eq:lab2prime-in}
  \end{align}
  Then $\compof{S}{\lab}'$ is a solution of $S$.
\end{theorem}
\begin{proof}
  \para{Preliminaries}
  By Theorem~\ref{thm:cut-partition-apx}, we have that $(T_{1},T_{2})$ is a partition of $S$.
  Consider a mapping $\compof{S}{\lab}' : \compof{S}{V} \rightarrow R$, defined as stated above.
  Even though none of the cases are over $\compof{S}{V}$,
  this defines $\compof{S}{\lab}'$ over all $\compof{S}{V}$, since $\compof{1}{V} \cup \compof{2}{V} = \compof{S}{V}$.
  No case is ever in conflict: by Lemma~\ref{lem:3},
  $\forall v \in \compof{1}{V} \cap \compof{2}{V}.~\compof{1}{\lab}(v) = \compof{2}{\lab}(v)$, so
  Equations~\eqref{eq:lab2prime-0} and~\eqref{eq:lab2prime-1} both apply for all shared nodes;
  by Lemma~\ref{lem:2} and the definition of fragments, if $v \in \compof{1}{\vin} \cap \compof{2}{\vin}$,
  then $\compof{1}{\inh}(v) = \compof{2}{\inh}(v) = \compof{S}{\inh}(v)$,
  so Equation~\eqref{eq:lab2prime-in} holds for any shared inputs.

  Our goal is to show that $\compof{S}{\lab}'$ is a solution for $S$ as stated in~\S\ref{sec:open-sol}.
  We proceed by considering an arbitrary node $u$, and show that, for each node subset $u$ could belong to
  ($u \notin \compof{S}{\vin}, u \in \compof{S}{\vin}, u \in \compof{S}{\vout}$),
  $\compof{S}{\lab}'(u)$ is a solution for $u$.

  \para{$u \notin \compof{S}{\vin}$ Case}
  Then we want to show that our mapping implies that
  $\compof{S}{\lab}'(u) = \init(u) \merge \Merge_{vu \in \compof{S}{E}} \transfer(vu, \compof{S}{\lab}'(v))$.
  We have two cases to consider here, depending on if $u$ is in a single fragment
  ($\compof{1}{V} \ominus \compof{2}{V}$, meaning \emph{either} $\compof{1}{V}$ \emph{or} $\compof{2}{V}$,
  the symmetric difference of $\compof{1}{V}$ and $\compof{2}{V}$),
  or whether $u \in (\compof{1}{V} \cap \compof{2}{V}) \setminus \compof{S}{\vin}$.

  \para{$u \in (\compof{1}{V} \ominus \compof{2}{V}) \setminus \compof{S}{\vin}$ Sub-Case}
  Suppose \WLOG{} that $u \in \compof{1}{V}$.
  Then since $T_{1}$ has a solution $\compof{1}{\lab}$, we have that
  $\compof{1}{\lab}(u) = \init(u) \merge \Merge_{vu \in \compof{1}{E}}\transfer(vu, \compof{1}{\lab}(v))$,
  since $u$ must be either a base node or an output node in $\compof{S}{\vout}$.

  In either such case, we then also know that $u$ has the same neighbors in $T_{1}$
  as in $S$, so
  $\{vu ~\vert~ vu \in \compof{1}{E}\} = \{vu ~\vert~ vu \in \compof{S}{E}\}$.
  By~\eqref{eq:lab2prime-0}, we have that $\compof{S}{\lab}'(u) = \compof{1}{\lab}(u)$ that for each neighbor $v$,
  $\compof{S}{\lab}'(v) = \compof{1}{\lab}(v)$, so we then can substitute $\compof{S}{\lab}'$ for $\compof{1}{\lab}$ and
  the set of neighbors in $\compof{S}{E}$ for the set of neighbors in $\compof{1}{E}$, giving
  $\compof{S}{\lab}'(u) = \init(u) \merge \Merge_{vu \in \compof{S}{E}}\transfer(vu, \compof{S}{\lab}'(v))$.
  Then this sub-case holds.

  \para{$u \in (\compof{1}{V} \cap \compof{2}{V}) \setminus \compof{S}{\vin}$ Sub-Case}
  By the shared input constraint,
  $(\compof{1}{V} \cap \compof{2}{V}) \setminus \compof{S}{\vin} = (\compof{1}{\vin} \setminus \compof{S}{\vin}) \cup (\compof{2}{\vin} \setminus \compof{S}{\vin})$.
  In other words, since $u \notin \compof{S}{\vin}$, it is an input-output node.

  Suppose \WLOG{} that $u \in \compof{1}{\vin} \setminus \compof{S}{\vin}$.
  Then $u \in \compof{2}{\vout}$ by the input-output constraints.
  Then since $T_{1}$ and $T_{S}$ have solutions, we have that
  \begin{align}
    \compof{1}{\lab}(u) &= \compof{1}{\inh}(u) \label{eq:lab0-input}\\
    \compof{2}{\lab}(u) &= \init(u) \merge \Merge_{vu \in \compof{2}{E}}\transfer(vu, \compof{2}{\lab}(v))%
    \label{eq:lab1-merge} \\
    \compof{2}{\lab}(u) &= \compof{2}{\outh}(u) \label{eq:lab1-output}
  \end{align}

  \noindent By these equations and the input-output constraints, we then have that
  $\compof{1}{\lab}(u) = \compof{1}{\inh}(u) = \compof{2}{\outh}(u) = \compof{2}{\lab}(u)$,
  so $\compof{1}{\lab}(u) = \compof{2}{\lab}(u)$.
  By~\eqref{eq:lab2prime-0} and~\eqref{eq:lab2prime-1}, we also have that
  $\compof{S}{\lab}'(u) = \compof{1}{\lab}(u) = \compof{2}{\lab}(u)$.

  Now we wish to show that $\compof{S}{\lab}'$ is a solution for $u$:
  since $u$ is not in $\compof{S}{\vin}$, we must show the non-input case
  $\compof{S}{\lab}'(u) = \init(u) \merge \Merge_{vu \in \compof{S}{E}}\transfer(vu, \compof{S}{\lab}'(v))$
  (we defer the output constraint $\compof{S}{\lab}'(u) = \compof{S}{\outh}(u)$ to the end of the proof).

  As above, we start by observing that $u$ has the same in-neighbors in $S$
  as in $T_{2}$, and that this encompasses all of its in-neighbors since
  $u \in \compof{1}{\vin}$, so it has no in-neighbors in $T_{1}$.
  Then $\{vu ~\vert~ vu \in \compof{2}{E}\} = \{vu ~\vert~ vu \in \compof{S}{E}\}$.

  Next, by~\eqref{eq:lab2prime-1}, we can substitute $\compof{S}{\lab}'$ for $\compof{2}{\lab}$
  in~\eqref{eq:lab1-merge}.
  By the reasoning above, we can also substitute the set of in-neighbors of $u$ in $\compof{S}{E}$
  for the set of in-neighbors of $u$ in $\compof{2}{E}$, leaving us with
  $\compof{S}{\lab}'(u) = \init(u) \merge \Merge_{vu \in \compof{S}{E}}\transfer(vu, \compof{S}{\lab}'(v))$.

  Then this case holds as well, and we have that $\compof{S}{\lab}'(u)$ is a solution
  for $u \notin \compof{S}{\vin}$.

  \para{$u \in \compof{S}{\vin}$ Case}
  Since $\compof{S}{\lab}'(u) = \compof{S}{\inh}(u)$ by~\eqref{eq:lab2prime-in}, this case immediately holds.

  \para{$u \in \compof{S}{\vout}$ Case}
  By the definition of a fragment,
  $\compof{1}{\lab}(u) = \compof{S}{\outh}(u)$ if $u \in \compof{1}{V}$ and
  $\compof{2}{\lab}(u) = \compof{S}{\outh}(u)$ if $u \in \compof{2}{V}$.
  Since $\compof{S}{\lab}'(u) = \compof{1}{\lab}(u)$ by~\eqref{eq:lab2prime-0} in the former case and
  $\compof{S}{\lab}'(u) = \compof{2}{\lab}(u)$ by~\eqref{eq:lab2prime-1} in the latter case,
  we have that $\compof{S}{\lab}'(u) = \compof{S}{\outh}(u)$, so this case holds.

  Then for all three cases, $\compof{S}{\lab}'$ is a solution for $S$.
\end{proof}

This concludes our proof of soundness: we now know that the solutions of the fragments
constitute a solution of the parent SRP\@.
We now also prove completeness, meaning
that any solution to the parent SRP is also a solution to the fragments,
so long as the fragments' inputs and outputs are annotated with assumptions and
guarantees that match the parent SRP's solution.
The form of the proof is also by solution cases, this time over the fragment solution.

\begin{theorem}[$\cut$ is Complete]\label{thm:completeness-apx}
  Let $S$ be an open SRP, and let $I$ be an interface over $S$.
  Let $\cut(S, I) = (T_{1}, T_{2})$.
  Assume $S$ has a unique solution $\compof{S}{\lab}$.
  Assume that $\forall uv \in \dom(I).~I(uv) = \compof{S}{\lab}(u)$.
  Consider the following two mappings $\compof{1}{\lab}' : \compof{1}{V} \rightarrow R$
  and $\compof{2}{\lab}' : \compof{2}{V} \rightarrow R$, defined such that:
  \begin{align*}
    \forall v \in \compof{1}{V}.~\compof{1}{\lab}'(v) &= \compof{S}{\lab}(v) \\
    \forall v \in \compof{2}{V}.~\compof{2}{\lab}'(v) &= \compof{S}{\lab}(v)
  \end{align*}
  Then $\compof{1}{\lab}'$ is a solution for $T_{1}$ and
  $\compof{2}{\lab}'$ is a solution for $T_{2}$.
\end{theorem}
\begin{proof}
  By Theorem~\ref{thm:cut-partition-apx}, we have that $(T_{1},T_{2})$ is a partition of $S$.
  Furthermore, by the definition of $\cut$ and the assumption that every cut edge is annotated with
  the solution in $S$, we have the following equalities on $T_{1}$ and $T_{2}$'s inputs and outputs:
  \begin{align}
    \forall u \in \compof{1}{\vin}.\compof{1}{\inh}(u) &= \compof{S}{\lab}(u)\label{eq:comp1} \\
    \forall u \in \compof{2}{\vin}.\compof{2}{\inh}(u) &= \compof{S}{\lab}(u)\label{eq:comp2} \\
    \forall u \in \compof{1}{\vout}.\compof{1}{\outh}(u) &= \compof{S}{\lab}(u)\label{eq:comp3} \\
    \forall u \in \compof{2}{\vout}.\compof{2}{\outh}(u) &= \compof{S}{\lab}(u)\label{eq:comp4}
  \end{align}

  As the two cases are symmetric, \WLOG, we proceed by considering an arbitrary node $u$ in $T_{1}$.
  Then we have three cases to show, based on the three cases of $\compof{1}{\lab}'(u)$.

  \para{$u \notin \compof{1}{\vin}$ Case}
  By the fragment constraints, if $u \in \compof{S}{\vin}$ then if $u \in \compof{1}{V}$, then $u \in \compof{1}{\vin}$.
  Then by the contrapositive, if $u \notin \compof{1}{\vin}$, then $u \notin \compof{S}{\vin}$.
  Then $\compof{S}{\lab}(u) = \init(u) \merge \Merge_{vu \in \compof{S}{E}}\transfer(vu, \compof{S}{\lab}(v))$.
  By the fact that $T_{1}$ is a fragment,
  $\{vu ~\vert~ vu \in \compof{1}{E}\} = \{vu ~\vert~ vu \in \compof{S}{E}\}$.
  Then, by our definition of $\compof{1}{\lab}'(u)$, we can substitute $\compof{1}{\lab}'(u)$ for $\compof{S}{\lab}(u)$ to
  obtain:
  $\compof{1}{\lab}'(u) = \init(u) \merge \Merge_{vu \in \compof{1}{E}}\transfer(vu, \compof{1}{\lab}'(v))$.
  Then this case holds for $u$.

  \para{$u \in \compof{1}{\vin}$ Case}
  Then by~\eqref{eq:comp1}, we have $\compof{S}{\lab}(u) = \compof{1}{\inh}(u)$.
  Then by substitution, we have $\compof{1}{\lab}'(u) = \compof{1}{\inh}(u)$.
  Then this case holds for $u$.

  \para{$u \in \compof{1}{\vout}$ Case}
  Then by~\eqref{eq:comp3}, we have $\compof{S}{\lab}(u) = \compof{1}{\outh}(u)$.
  Then by substitution, we have $\compof{1}{\lab}'(u) = \compof{1}{\outh}(u)$.
  Then this case holds for $u$.

  Then $T_{1}$ has a solution $\compof{1}{\lab}'$.
  By a symmetric proof using~\eqref{eq:comp2} and~\eqref{eq:comp4},
  $T_{2}$ has a solution $\compof{2}{\lab}'$.
\end{proof}

An important corollary of our theorem of soundness is that, since the solutions
of the fragments are a solution to the parent SRP, any property over solutions
that holds on the fragments will also hold on the parent SRP\@.

\begin{corollary}[$\cut$ Preserves Properties]\label{cor:properties-apx}
  Let $S$ be an open SRP, and let $I$ be an interface over $S$.
  Let $\cut(S, I) = (T_{1}, T_{2})$.
  Let $P_{1}, P_{2}$ be formulas such that
  $P_{1} = \forall v \in \compof{1}{V}.~Q(v)$ and
  $P_{2} = \forall v \in \compof{2}{V}.~Q(v)$, where $Q$ is a predicate on $\lab(v)$.
  Assume $S$ has a unique solution $\compof{S}{\lab}$,
  and that $T_{1}$ has a solution $\compof{1}{\lab}$
  and $T_{2}$ has a solution $\compof{2}{\lab}$.
  Then if $P_{1}$ holds on $T_{1}$ and $P_{2}$ holds on $T_{2}$,
  $P_{1} \wedge P_{2}$ holds on $S$.
\end{corollary}
\begin{proof}
  By Theorem~\ref{thm:soundness-apx},
  $\forall u \in \compof{1}{V}.~\compof{1}{\lab}(u) = \compof{S}{\lab}(u)$
  and $\forall u \in \compof{2}{V}.~\compof{2}{\lab}(u) = \compof{S}{\lab}(u)$.
  Assume $P_{1}$ holds on $T_{1}$ and $P_{2}$ holds on $T_{2}$.
  Consider \WLOG{} a node $u$ in $\compof{1}{V}$.
  Then $Q(u)$ holds in $T_{1}$.
  Since $\compof{1}{\lab}(u) = \compof{S}{\lab}(u)$, $Q(u)$ holds in $S$ as well.
  Then since $\compof{1}{V} \cup \compof{2}{V} = \compof{S}{V}$,
  $\forall v \in \compof{S}{V}.~Q(v)$, and therefore $P_{1} \wedge P_{2}$  holds on $S$
\end{proof}

%
\end{document}